\documentclass{aa}  
 
\usepackage{graphicx}
\usepackage{txfonts}

\usepackage{hyperref}
\hypersetup{
    colorlinks = True,
    citecolor=blue,
    linkcolor = cyan,
}

\usepackage{placeins}

\DeclareUnicodeCharacter{2212}{-}
\usepackage{xcolor} 
\usepackage{tabu}

\usepackage{booktabs}  
\usepackage{multirow} 
\usepackage{xspace}
\usepackage{float}
\usepackage{stfloats}

\begin{document} 

\newcommand{\ngc}{NGC\,1365\xspace}

\let\oldAA\AA
\renewcommand{\AA}{\text{\oldAA}\xspace}
\let\oldarcsec\arcsec
\renewcommand{\arcsec}{\text{\oldarcsec}\xspace}

\newcommand{\red}[1]{\textcolor{red}{#1}}
\newcommand{\nevA}{[\ion{Ne}{V}]$\lambda14\mu\mathrm{m}$\xspace}
\newcommand{\nevB}{[\ion{Ne}{V}]$\lambda24\mu\mathrm{m}$\xspace}
\newcommand{\arvA}{[\ion{Ar}{V}]$\lambda8\mu\mathrm{m}$\xspace}
\newcommand{\arvB}{[\ion{Ar}{V}]$\lambda13\mu\mathrm{m}$\xspace}
\newcommand{\mgvA}{[\ion{Mg}{V}]$\lambda5\mu\mathrm{m}$\xspace}
\newcommand{\mgvB}{[\ion{Mg}{V}]$\lambda13\mu\mathrm{m}$\xspace}

\newcommand{\htwo}{H$_2$~$0-0$~S(1)\xspace}
\newcommand{\oiii}{[\ion{O}{III}]$\lambda5007\AA$\xspace}
\newcommand{\fevii}{[\ion{Fe}{VII}]$\lambda6087\AA$\xspace}
\newcommand{\siiALL}{[\ion{S}{II}]$\lambda\lambda6716,6731$\xspace}
\newcommand{\siia}{[\ion{S}{II}]$\lambda6716$\xspace}
\newcommand{\siib}{[\ion{S}{II}]$\lambda6731$\xspace}

\newcommand{\neii}{[\ion{Ne}{II}]$\lambda13\mu\mathrm{m}$\xspace}
\newcommand{\co}{CO(3-2)\xspace}

\newcommand{\msun}{\ensuremath{\mathrm{M_\odot}}\xspace}
\newcommand{\MOKA}{\ensuremath{\mathrm{MOKA^{3D}}}\xspace}

\title{MIRACLE}
\subtitle{II. Unveiling the multiphase gas interplay in the circumnuclear region of NGC~1365 via multicloud modeling}

\titlerunning{Unveiling the multiphase gas interplay in the circumnuclear region of NGC 1365 via multicloud modeling}
\authorrunning{Ceci et al.}

\author{
    M.~Ceci \inst{1,2}
\and C.~Marconcini\inst{1,2}
\and A.~Marconi \inst{1, 2}
\and A.~Feltre \inst{2}
\and I.~Lamperti \inst{1,2}    
\and F.~Belfiore \inst{2, 3}
\and E.~Bertola \inst{2}
\and C.~Bracci \inst{1,2}
\and S.~Carniani \inst{4}
\and E.~Cataldi \inst{1,2}
\and G.~Cresci \inst{2}
\and Q.~D'Amato\inst{2}
\and J.~Fritz \inst{5}
\and M.~Ginolfi \inst{1,2}
\and E.~Hatziminaoglou \inst{3,6,7}
\and M.~Hirschmann \inst{8}
\and M.~Mingozzi \inst{9}
\and B.~Moreschini \inst{1,2}
\and F.~Mannucci \inst{2}
\and G.~Sabatini \inst{2}
\and F.~Salvestrini \inst{10,11}
\and M.~Scialpi \inst{1,2,12}
\and G.~Tozzi \inst{13}
\and L.~Ulivi \inst{1,2,12}
\and G.~Venturi \inst{4,2}
\and A.~Vidal-Garc\'ia \inst{14}
\and C.~Vignali \inst{15,16}
\and M.~V.~Zanchettin \inst{2}
    }
 
\institute{Università di Firenze, Dipartimento di Fisica e Astronomia, via G. Sansone 1, 50019 Sesto F.no, Firenze, Italy\\\email{matteo.ceci@unifi.it}
\and  
INAF - Osservatorio Astrofisico di Arcetri, Largo E. Fermi 5, I-50125 Firenze, Italy
\and   
European Southern Observatory, Karl-Schwarzschild Straße 2, D-85748 Garching bei München, Germany
\and   
Scuola Normale Superiore, Piazza dei Cavalieri 7, 56126 Pisa, Italy
\and  
Instituto de Radioastronomía y Astrofísica, Universidad Nacional Autónoma de México, Morelia, Michoacán 58089, Mexico
\and   
Instituto de Astrof\'{i}sica de Canarias, 38205 La Laguna, Tenerife, Spain 
\and   
Departamento de Astrof\'{i}sica, Universidad de La Laguna, 38206 La Laguna, Tenerife, Spain
\and   
Institute of Physics, GALSPEC laboratory, EPFL, Observatory of Sauverny, Chemin Pegasi 51, 1290 Versoix, Switzerland
\and   
AURA for ESA, Space Telescope Science Institute, 3700 San Martin Drive, Baltimore, MD 21218, USA
\and   
INAF, Osservatorio Astronomico di Trieste, Via Tiepolo 11, I-34131 Trieste, Italy 
\and   
IFPU - Institute for Fundamental Physics of the Universe, via Beirut 2, I-34151 Trieste, Italy
\and   
University of Trento, Via Sommarive 14, Trento, I-38123, Italy
\and  
Max-Planck-Institut für Extraterrestrische Physik (MPE), Gießenbachstr. 1, D-85748 Garching, Germany
\and   
Observatorio Astronómico Nacional, C/ Alfonso XII 3, 28014 Madrid, Spain
\and   
Dipartimento di Fisica e Astronomia, Alma Mater Studiorum, Università degli Studi di Bologna, Via Gobetti 93/2, 40129 Bologna, Italy
\and  
INAF–Osservatorio di Astrofisica e Scienza dello Spazio di Bologna, Via Gobetti 93/3, 40129 Bologna, Italy
}

\date{Received 07 15, 2025; accepted 22 10, 2025}

\abstract
{
We present a multiphase analysis of the gas in the circumnuclear region ($\sim$ 0.9$\times$0.9 kpc$^2$) of the nearby barred Seyfert 1.8 galaxy NGC 1365, observed as part of the Mid-IR Activity of Circumnuclear Line Emission (MIRACLE) program. Specifically, we combined spatially resolved spectroscopic data from JWST/MIRI, VLT/MUSE, and ALMA to provide a multiphase characterization of the ionized atomic and the warm and cold molecular gas phases.
MIRI data enabled the detection of more than 40 mid-IR emission lines from ionized or warm molecular gas. Moment maps show that both cold and warm molecular gas trace the circumnuclear ring, following the rotation of the stellar disk. 
The ionized gas exhibits flux distributions and kinematics that vary depending on the ionization potential (IP). Low-IP species ($\leq$~25 eV) mainly trace the rotating disk, while higher-IP species (up to $\sim$120 eV) trace the outflowing gas. 
Both \oiii and \nevA trace the nuclear outflow cone toward the southeast. In addition, the \nevA line traces the counter-cone of the outflow to the northwest, which is obscured in the optical at these circumnuclear scales, and is thus undetected in \oiii.
Unlike optical diagnostics, spatially resolved mid-IR diagnostics reveal the key role of the active galactic nucleus (AGN) as the source of gas ionization in the central region.
We derived the electron density from the \nevB/\nevA line ratio, finding a median value of (750$\pm$440)~cm$^{-3}$, consistent with previous estimates obtained from the optical [\ion{S}{II}] doublet.
Lastly, we applied, for the first time, a fully self-consistent combination of state-of-the-art photoionization and kinematic models (HOMERUN+\MOKA) to estimate the intrinsic physical outflow properties, kinematics, and energetics - overcoming the limitations of classical methods based on oversimplified assumptions. Exploiting the unprecedented synergy between JWST/MIRI and VLT/MUSE, HOMERUN allows us to simultaneously reproduce the fluxes of over 60 emission lines spanning from the optical to the mid-IR. This unique approach enables us to disentangle the physical conditions of AGN- and star formation-dominated components and robustly estimate the mass of the outflowing gas and other physical properties.
}

\maketitle
%
\section{Introduction}
Galactic outflows are known to influence the evolution of their host galaxies by removing gas, enriching the circumngalactic medium, and suppressing star formation (SF) in the galaxy \citep{Veilleux2005, Fabian2012, CresciMaiolino2018,Harrison2024}. While powerful outflows are commonly observed in luminous active galactic nuclei (AGNs) at redshifts $z=1 - 3$ \citep[e.g.,][]{CanoDiaz2012, Fiore2017}, similar signatures are also detected in local, less powerful AGN-hosted galaxies, where the high spatial resolution enabled by their vicinity allows for detailed studies of their physical and kinematic properties \citep{Cresci2015, Venturi2017,Venturi_2018, Cicone2018, Mingozzi2019,Fluetsch2019, Lutz2020}. These local outflows span multiple gas phases, from ionized atomic to molecular, and show broad, blueshifted components in emission and/or absorption lines, consistent with being driven by AGN activity.

Ionized outflows, often traced by strong optical emission lines such as \oiii, exhibit velocities ranging from several hundred to a few thousand kilometers per second. They are composed of warm gas (T~=~$10^4$~K, n$_e$=$10^2-10^4$cm$^{-3}$) that expands from the nuclear region out to galactic scales, potentially expelling material from their host galaxy \citep[e.g.,][]{Harrison2014,harrison16, Woo2016, Fiore2017, Venturi_2018, ForsterSchreiber2019,Kakkad2020, Venturi2021_turmoil, Cresci2023,Speranza2024, Marconcini2025_nat}. 
However, the impact of ionized winds on SF remains debated \citep[e.g.,][]{harrison16, Harrison2024}, as their coupling efficiency with the ambient interstellar medium (ISM) appears to be low. Indeed, the energy and momentum carried by the ionized gas phase are typically subdominant when compared to those of other gas phases \citep[e.g.,][]{Zubovas2017, Combes2017, Fluetsch2019,Fluetsch2021,Mulcahey2022, Belli2024}. In the context of these findings, the ionized gas component may only trace the larger-scale, less massive regions of multiphase outflows, which are dominated in mass by colder molecular or atomic gas on less extended scales \citep[e.g.,][]{RamosAlmeida2022, Audibert2023,Venturi2023}.

Cold molecular outflows represent a colder phase of gas (T~=~$10-10^2$~K, n$_e~\ge10^3$ cm$^{-3}$), frequently observed via CO emission lines. Compared to the ionized phase, CO-traced outflows are typically slower, with velocities of a few hundred kilometers per second, but often carry substantial mass, suggesting they may play a crucial role in regulating SF \citep{Feruglio2010, Fiore2017, GarciaBurillo2019,Lutz2020,Fluetsch2019,Fluetsch2021}.
In addition to the cold component, warm molecular outflows (T~=~$10^2-10^3$~K) are typically traced by H$_2$ roto-vibrational transitions in the near- and mid-infrared (IR). H$_2$ emission is particularly prominent in shocked regions, making it a sensitive tracer of warm molecular gas entrained in outflows \citep[][]{Hill2014, Richings2018, Richings2018b, Riffel2020, Wright2023}.

\ngc, the subject of this work, is a great laboratory in which to study in detail multiphase AGN feedback. This dusty barred spiral galaxy (SB(s)b; \citealt{deVaucouleurs1991}) is located in the Fornax cluster \citep{Jones1980} at a distance of 19.57 Mpc \citep{Jacobs09,Anand2021} (1\arcsec $\sim$ 95 pc).
With a redshift of z = 0.005457 \citep{Bureau1996}, \ngc is classified as a Seyfert 1.8 galaxy \citep{Veron2006}, with signatures of both AGN activity and SF in its central regions. 

The AGN-driven ionized outflow in \ngc is observed via extended biconical \oiii emission spanning $\sim$2.5 kpc, with line-of-sight velocities of up to $\pm$170 km s$^{-1}$ \citep{Venturi_2018}. The southeastern (SE) cone, approaching the observer, and the receding northwestern (NW) cone -- partially obscured by the galactic disk -- are dominated by AGN ionization as traced by diagnostic diagrams \citep{Baldwin1981, Veilleux1987, Kewley+06}. High-resolution X-ray observations revealed a fast, highly ionized blueshifted nuclear (spatially unresolved) wind in absorption with velocities of $\sim$ 3000 km s$^{-1}$, further supporting the scenario of an AGN-driven wind \citep{Risaliti2005,Braito2014}.

\ngc has a SF rate of 16.9 M$_{\odot}$ yr$^{-1}$ \citep{Lee2022}, mainly within a $\sim$2~kpc radius circumnuclear ring identified across optical, radio, and IR wavelengths \citep{Kristen1997,Forbes1998,Alonso-Herrero2012}. The ring is associated with the inner Lindblad resonance \citep{Lindblad1996} and shows noncircular bar-driven gas motions \citep{Teuben1986, Sanchez2009}, associated with
SF-driven ionization \citep{Sharp2010, Venturi_2018} and large reservoirs of molecular gas \citep[$\sim10^9$ \msun;][]{Sakamoto2007, Gao2021}. 
Observational campaigns from the PHANGS survey (``Physics at High Angular resolution in Nearby GalaxieS'', \citealt{Lee2023})
conducted with the \textit{James Webb} Space Telescope (JWST) and complemented by the Atacama Large Millimeter/submillimeter Array (ALMA; \citealt{Wootten2009})
have revealed that young clusters embedded in the ring drive both ionization of the surrounding medium and localized modifications to the molecular gas phase, where feedback-driven heating and dissociation decrease CO excitation and increase [C I]/CO abundance ratios in their immediate vicinity \citep{Liu2023, Schinnerer2023}.

Despite hints of a nuclear radio jet in \ngc \citep{Sandqvist1995}, later studies by \citet{Stevens1999} found no conclusive evidence of its presence. Instead, radio emission is primarily attributed to the elongated star-forming ring rather than a jet.

JWST is transforming our understanding of the outflow properties in the local Universe by providing an unprecedented view of the nuclear dusty regions of AGNs \citep[e.g.,][]{ HermosaMunoz+2024_6240, Perna2024, Zhang2024, GarciaBernete2024, Davies2024,Ulivi2025,Ceci_2025}.
This revolution is largely driven by the unique capabilities of the JWST Mid-Infrared Instrument \citep[MIRI;][]{Wright15,Wright2023}, including its high spatial resolution across a wide range of wavelengths (from 5 $\mu$m to 28 $\mu$m), enabling detailed studies of the energetics and kinematics of gas flows in obscured nuclear regions of AGNs. The wide IR spectral range of JWST spanning many atomic and molecular gas transitions, combined with its spatial and spectral resolution, provides essential insights into the multiphase circumnuclear medium. These unprecedented capabilities enable multiphase, co-spatial analyses when combined with complementary data from ground-based facilities, such as the Multi Unit Spectroscopic Explorer (MUSE) at the ESO Very Large Telescope (VLT; \citealt{Bacon2010}) and ALMA, transforming our understanding of feedback mechanisms in local AGNs.

For NGC 1365, we combined data from the Medium-Resolution Spectrometer (MRS; \citealt{Rieke2015,Labiano_21}) of MIRI with integral-field optical spectroscopy from VLT/MUSE and millimeter observations from ALMA. This multiwavelength approach enables us to trace the multiphase gas across a wide range of ionization states and densities. In particular, it allows for a detailed characterization of the kinematics, morphology, and energetics of the circumnuclear environment in NGC 1365, ultimately contributing to a deeper understanding of AGN feedback and its impact on galaxy evolution.

This paper is organized as follows. 
In Section \ref{sec: Data reduction and emission line fitting}, we introduce the observations and the spectroscopic analysis of the MIRI data. 
In Section \ref{sec results and discussion}, we discuss our results through the characterization of the ISM properties and the spatially resolved kinematics of the ionized and molecular gas in the central region of NGC 1365. Finally, in Section \ref{sec conclusion}, we summarize our results. 
For all the maps shown in this work, north is up and east is to the left. 

\begin{figure*}[!]
    \centering
    \includegraphics[width=.9\linewidth]{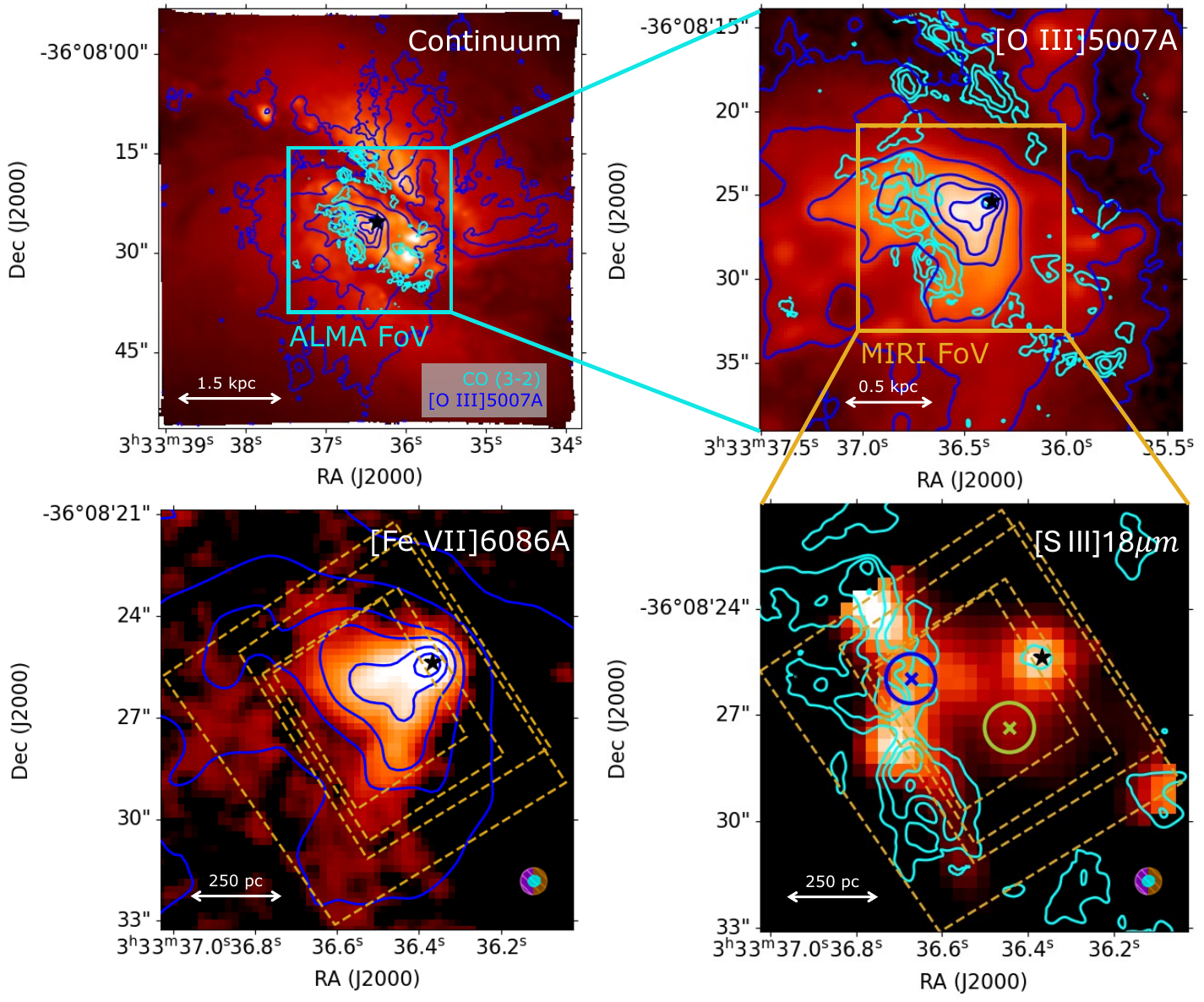}
    \caption{
    MIRI, MUSE, and ALMA observations of \ngc.
    \textit{Upper left:} Continuum map from MUSE data obtained collapsing the data in the wavelength range 5200-5800 \AA. 
    \textit{Upper right:} \oiii flux map from MUSE data in the ALMA FoV.
    \textit{Lower left:} \fevii flux map from MUSE data in the MIRI Ch4 FoV. 
    \textit{Lower right:} [S III]18.51$\mu$m flux map from MIRI data. Dashed orange rectangles represent the MIRI MRS channels FoV. Blue and green circles of radius of 0.7\arcsec represent the regions from which we extracted the spectra shown in Fig.~\ref{fig: spectra}.
    Cyan and blue contours represent arbitrary levels of CO(3-2) flux from ALMA and \oiii from MUSE, respectively.
   The star marks the position of the nucleus based on the ALMA data (see Section \ref{app:ALMA_data}). In the lower panels, the violet and orange circles represent the MUSE and MIRI PSF, respectively. The ALMA beam is shown as a cyan oval.
    }
    \label{fig: comparison fov}
\end{figure*}

\section{Overview of observations and data analysis}\label{sec: Data reduction and emission line fitting}

\ngc was observed on 8 December 2024 as part of the ``Mid-IR Activity of Circumnuclear Line Emission'' (MIRACLE; JWST GO program 6138; Co-PIs: C. Marconcini and A. Feltre), aimed at tracing the mid-IR emission by exploiting MIRI/MRS data of the circumnuclear region of local AGNs in the 5-28 $\mu$m wavelength range.
The MRS mode of MIRI covers a total wavelength range of 4.9–27.9 $\mu$m, divided into four integral field units (IFUs), also referred to as channels (Ch1, Ch2, Ch3, and Ch4, hereafter), each further subdivided into three bands (SHORT, MEDIUM, and LONG). These channels cover slightly different fields of view (FoVs), from 3.2\arcsec\ $\times$ 3.7\arcsec in Ch1 to 6.6\arcsec\ $\times$ 7.7\arcsec in Ch4, at varying pixel sizes (from 0.13\arcsec in Ch1 to 0.35\arcsec in Ch4) and resolving powers \citep[from $\sim$3700 to $\sim$1500; see e.g.,][]{Labiano_21, Argyriou2023}. 
As shown in \citet{Law2023}, the average FWHM of the MIRI/MRS PSF ranges from $\sim$0.4\arcsec in Ch1 to $\sim$0.9\arcsec in Ch4.
We summarize the data reduction in Appendix~\ref{app: data MIRI}, referring to \cite{MIRACLE_NGC424} for a detailed description. As a result of this process, the pipeline produced 12 datacubes, one for each sub-band. 

\begin{figure*}[!]
    \centering
    \includegraphics[width=.95\linewidth]{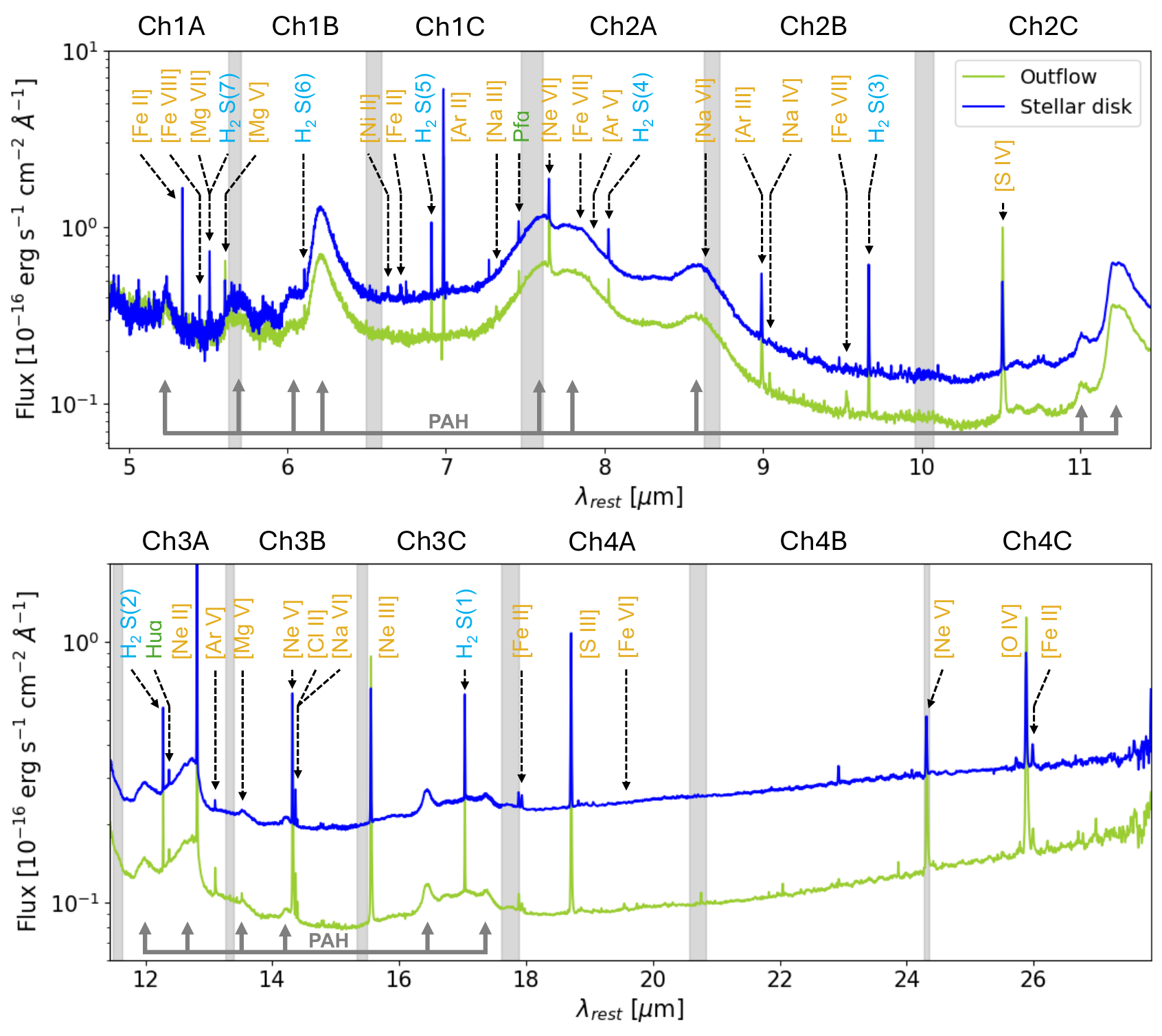}
    \caption{
    Integrated spectra of NGC 1365 from MIRI MRS data. The blue and green curves represent the integrated spectra extracted from the 0.7\arcsec radius apertures marked by the circles in the lower right panel of Fig.~\ref{fig: comparison fov}. These apertures sample regions dominated by the stellar disk and outflow emission, respectively.
    Detected emission lines are marked with vertical lines: ionized gas emission lines are labeled in yellow, H$_2$ rotational lines in cyan, and H I recombination lines in green. We annotate the main PAH features with gray arrows.
    The names of each MIRI MRS sub-channel are indicated, and the gray regions represent the overlapping spectral ranges of two adjacent sub-channels.
    }

    \label{fig: spectra}
\end{figure*}

The science observations targeted the nuclear region of \ngc but were deliberately offset to include the inner part of the SE approaching outflow. As is shown in the lower panels of Fig.~\ref{fig: comparison fov}, the different MIRI FoVs of each channel are not exactly centered on the nucleus, but shifted toward the SE to better capture the outflow region. This choice was motivated by the primary goal of our MIRACLE project of characterizing multiphase gas outflows and nuclear properties.
In the lower right panel of Fig.~\ref{fig: comparison fov}, we show the nuclear region of \ngc in the MIRI FoV. We extracted integrated spectra from the 0.7\arcsec radius circular apertures, and we present them in Fig.~\ref{fig: spectra}.
The green (blue) spectrum is extracted from the circular region shown in Fig.~\ref{fig: comparison fov} where the outflow (stellar disk) emission is dominant, as is shown by the flux maps in Fig. \ref{fig: moment maps}. Note that the stellar disk spectrum is contaminated by the outflow emission, as is indicated by the presence of high-excitation lines. We detected more than 40 emission lines, with a signal-to-noise ratio (S/N)~$\ge$~5, in the spectral range covered by MIRI, mostly tracing ionized gas, with an ionization potential (IP) -- defined as the energy required to create the relevant species -- ranging from a few electronvolts to more than 100 eV (see Table \ref{tab: table list emission lines}). Additionally, we also detected two hydrogen recombination lines; namely, Pf$\alpha$ and Hu$\alpha$ (i.e., H I (6−5) at 7.46 $\mu$m and H I (7−6) at 12.37 $\mu$m, respectively) and seven H$_2$ pure-rotational transitions, from 0-0 S(7) to 0-0 S(1).

As is shown in Fig. \ref{fig: spectra}, the MIRI spectra also exhibit at least 15 bright PAH features in the wavelength range between 5~-~18~$\mu$m, which we will investigate in a dedicated forthcoming paper.
To obtain these spectra, for the first time we combined the emission from all 12 MIRI datacubes, taking into account the different pixel sizes and ensuring a proper flux conservation among different bands by applying a scaling factor to align the flux levels between adjacent bands. A detailed description of this novel procedure is presented in the Appendix~\ref{app. MIRI corrections}. 
This algorithm, written in Python, is available for download\footnote{\url{https://github.com/matteo-ceci/JWST-MIRI_stitching_code}}.

In Appendix~\ref{app miri emission line fitting}, we analyzed the MIRI MRS datacubes using a customized python script to enhance the S/N with spatial smoothing, subtract the local continuum around all the brightest emission lines and then fit them spaxel by spaxel. The fit emission lines are listed in Table \ref{tab: table list emission lines}.

The MUSE IFU observations of \ngc were obtained on the 12th October 2014, under program 094.B-0321(A) (PI A. Marconi) and are part of the ``Measuring AGN Under MUSE'' (MAGNUM) survey \citep{Cresci2015, Venturi2017, Venturi_2018, Mingozzi2019, Marconcini2023, Marconcini2025_nat}. We applied the same data reduction pipeline that has been extensively applied in previous works \citep[e.g.,][]{Venturi_2018, Mingozzi2019}, and which has been validated and described in detail therein.
To trace the cold molecular gas, we used archival ALMA 12-m Band 7 observations (program 2016.1.00296.S, PI F.Combes; \citealt{Combes_2019}) of the CO J=$3\rightarrow2$ transition at 345.796GHz (rest-frame).
The observation description, data reduction and emission line fitting procedure of the MUSE and ALMA data can be found in Appendices \ref{app: data MUSE}-\ref{app MUSE emission line fitting} and \ref{app:ALMA_data}-\ref{app ALMA fitting}, respectively.

\section{Results}\label{sec results and discussion}

\subsection{Multiphase gas kinematics} \label{sec Multi-phase gas kinematics}
\begin{figure*}[!t]
    \centering
    \includegraphics[width=.9\linewidth]{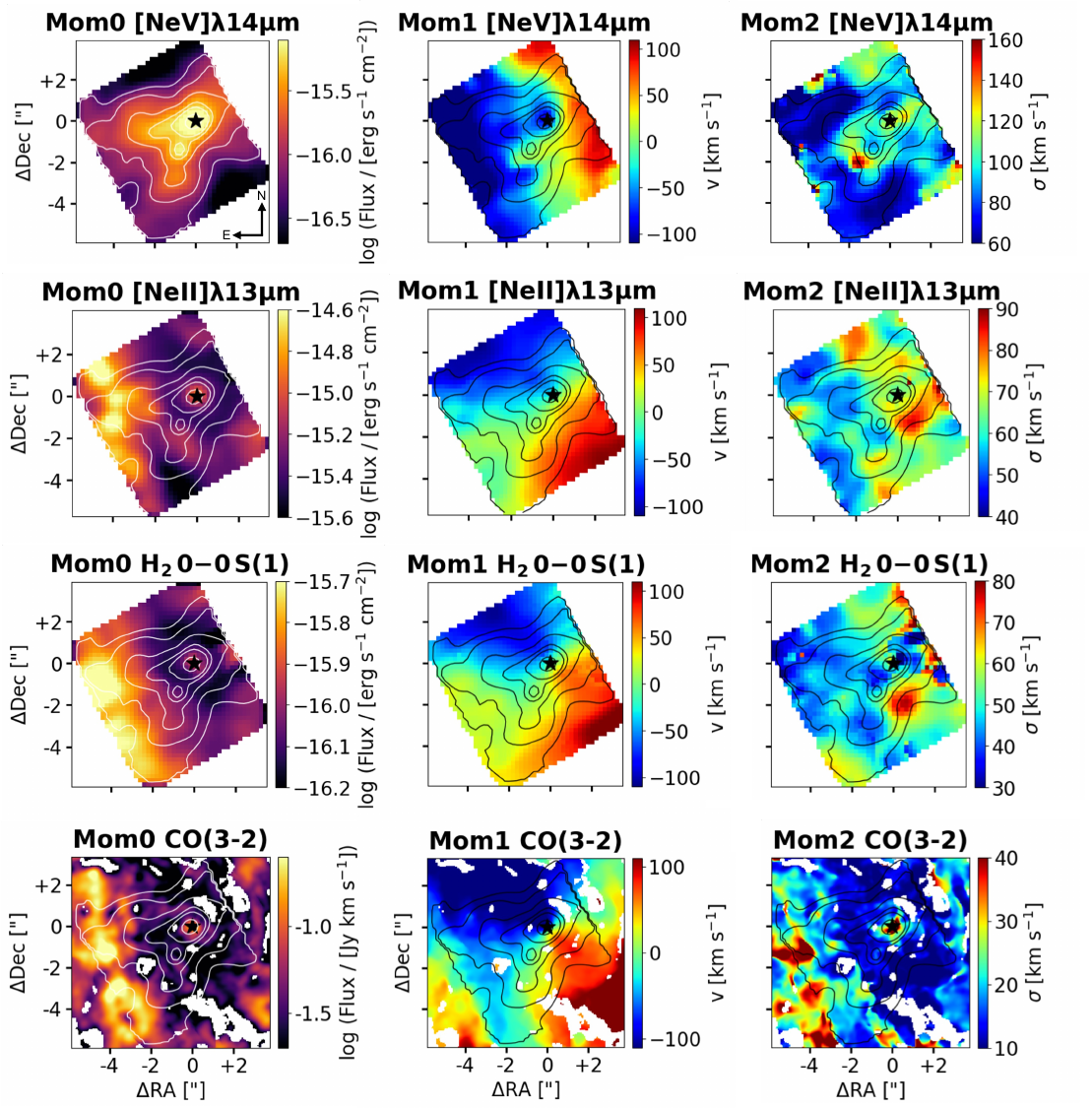}
    
    \caption{
    Moment maps of ionized and molecular gas emission in \ngc, tracing both the rotating disk and the outflowing gas components. From top to bottom: \nevA, \neii, \htwo, and \co moment maps.
    From left to right, we show the flux, the line-of-sight velocity (LOSV), and the velocity dispersion map. The contours represent arbitrary flux levels of \nevA emission. The star marks the nucleus position. The maps size is 870$\times$910 pc$^2$. Spaxels with S/N < 5 are masked.}
    \label{fig: moment maps}
\end{figure*}

For the kinematic analysis, we focused on emission lines within MIRI Ch3 (0.2\arcsec/pixel in a 6.6\arcsec~$\times$~7.7\arcsec FoV, corresponding to 19~pc/pixel over a 620~$\times$~730~pc$^2$ FoV) because of the good balance between spatial resolution and field coverage. Moreover, the Ch3 emission lines are among the brightest detected in our data, allowing us to present spatially resolved results representative of the ionized and molecular emission, with sufficient S/N across the entire FoV. Although in this analysis we focus on the FoV of Ch3, which also includes a fraction of disk gas illuminated by the AGN and not part of the outflow, it would be more accurate to refer to this region as the ionization cone rather than calling it outflow. However, we will assume in the following that all the gas in the ionization cone is in outflow.

Figure~\ref{fig: moment maps} shows the moment maps of \nevA, \neii, \htwo, and \co, obtained from the emission line fitting routines presented in Appendices \ref{app miri emission line fitting} and \ref{app ALMA fitting}. 
The \nevA seems to follow the direction of the ionized outflow; that is, the SE-NW direction (\citealt{Venturi_2018}), as expected from species with high-IP (see Table \ref{tab: table list emission lines}). Its velocity field is also consistent with the \oiii kinematics, which is blueshifted in the SE and redshifted in the NW. In the SE (NW) cone, we estimate gas projected velocities up to -120 km~s$^{-1}$ (+100 km~s$^{-1}$). 
The velocity dispersion map of \nevA shows a donut-shaped structure in SE direction of the nucleus (highlighted with contours in Fig. \ref{fig: velocity channels}), at a projected distance of $\sim$ 2.5\arcsec ($\sim$ 240 pc), with a peak of $\sigma$ $\sim$ 170 km s$^{-1}$.

The lower-IP \neii emission line has a different morphology with respect to the highly ionized gas traced by the \nevA, suggesting that it traces a different gas component. Indeed, the \neii morphology resembles the stellar continuum emission in the circumnuclear ring (whose ALMA contours are highlighted in Fig. \ref{fig: comparison fov}; see also \citealp{Liu2023}), as we can see in Fig. \ref{fig: moment maps}. Moreover, the moment-0 map reveals bright clumps and lanes in the left part of the FoV that match remarkably well the features visible in H$\alpha$ map of \citet{Venturi_2018}. This supports the interpretation that the \neii line predominantly traces the circumnuclear ring.
This includes the elongated structure $\sim$4\arcsec eastward of the nucleus, which is part of the so-called “mid-east” region -- associated with the Southern Arm, a stream of gas flowing in from the southeast along the bar -- and is interpreted as material moving downstream within the circumnuclear ring \citep{Liu2023}.
The fact that \neii traces the disk component is also confirmed by the velocity gradient of \neii, whose direction, along the galaxy disk major axis, and magnitude ($\sim 120$ km s$^{-1}$) are consistent with those of the rotating stellar component (and H$\alpha$) observed on the same scales by \cite{Venturi_2018}.

The warm molecular gas component, which is traced by the H2 0-0 S(1) emission line, shares a similar morphology as that of the \neii line (Fig. \ref{fig: moment maps}). 
The warm molecular gas flux peaks at larger distances from the nucleus and appears more diffuse, closely resembling the CO(3–2) emission. In contrast, the ionized gas traced by \neii is concentrated in more compact regions. Since the spatial resolution at the wavelengths of the two lines is comparable, this difference in morphology reflects an intrinsic difference in the extent of the emitting regions.
In addition, we observe a concentration of warm molecular gas at the position of the nucleus, which could be explained by the presence of large amounts of dust (as seen in dust continuum from MIRI by \citealp{Liu2023}) shielding the molecules from the AGN radiation. 
The kinematics follows the same rotational pattern observed in \neii, with some deviations from a pure rotating disk. In particular, the NW side of the H$_2$ velocity map shows lower blueshifted velocities compared to the \neii velocity. On the opposite side, a deviation from rotation is observed in the redshifted region, which is co-spatial with an enhancement in velocity dispersion. This region, located 2\arcsec south of the nucleus, exhibits a peak in dispersion reaching 80 km s$^{-1}$ and appears as a distinct region.

The \co moment maps shown in Fig. \ref{fig: moment maps} are in agreement with the \htwo kinematics, which is tracing the galaxy rotating disk. A similar CO velocity field was reported by \citet{Liu2023}. This finding, which is consistent with the lack of any molecular outflow in the circumnuclear region \citep[see also][]{Combes_2019}, is in contrast to the CO (1-0) wind reported by \cite{Gao2021}. Their outflow interpretation was based on velocity residuals from a rotating disk model on larger spatial scales, which resemble those observed in H$\alpha$ and are more likely due to noncircular inflowing motions along the bar (see their Fig.~6).
Notably, the cold gas traced by the \co does not show the $\sigma$-enhanced region south of the nucleus as the H$_2$, although both emission lines map the molecular gas. 
This suggests that the elevated velocity dispersion in H$_2$ may trace noncircular motions, possibly related to local turbulence or streaming motions induced by the AGN, rather than a genuine outflow. In such an environment, the cold molecular gas traced by CO may not survive due to stronger dissociation or heating, leaving only the warm phase observable in this disturbed region.

The moment maps of the other emission lines listed in Table \ref{tab: table list emission lines} are shown in the \href{https://zenodo.org/records/17535125}{online material}, divided into molecular gas, recombination lines, and species with low ($<$~25 eV), medium ($>$~25 and $<$~54 eV), and high ($>$~54 eV) IP. The thresholds are based on the IPs of \ion{He}{I} ($\sim$25 eV) and \ion{He}{II} ($\sim$54 eV).
Note that while some lines appear to lack emission in the nuclear region of the galaxy in the moment maps, this is because the spectra are dominated by strong continuum emission rather than a genuine absence of emission.
Grouping such emission lines by IP or gas phase leads to interesting typical patterns.
All the H$_2$ lines share kinematic features similar to those of \htwo in Fig.~\ref{fig: moment maps}, with a rotational disk kinematics and a high-$\sigma$ blob to the south of the nucleus. 
The low-IP lines, as well as the recombination lines, show a velocity gradient consistent with a rotating disk, with an amplitude of $\sim$ 110 km s$^{-1}$, like \neii. The high-$\sigma$ region observed in H$_2$ is not present in all the low-IP and recombination lines.
From the flux maps, we observed that these species are located mainly in the mid-east region \citep{Liu2023}.
The moment maps of the high-IP species are similar to those of \nevA (see Fig.~\ref{fig: moment maps}). In particular, the donut-shaped structure is well defined in the velocity dispersion maps, although in the Ch4 the spatial resolution is lower and the shape is not well recognizable. 
The medium-IP lines have intermediate properties compared to the previous two groups: the velocity gradient is similar to that of the low-IP lines, but the velocity dispersion maps show a high-velocity dispersion region in correspondence of the donut shape.

\subsection{Velocity channel maps} \label{sec Velocity channel maps}
\begin{figure*}
    \centering
    \includegraphics[width=0.9\linewidth]{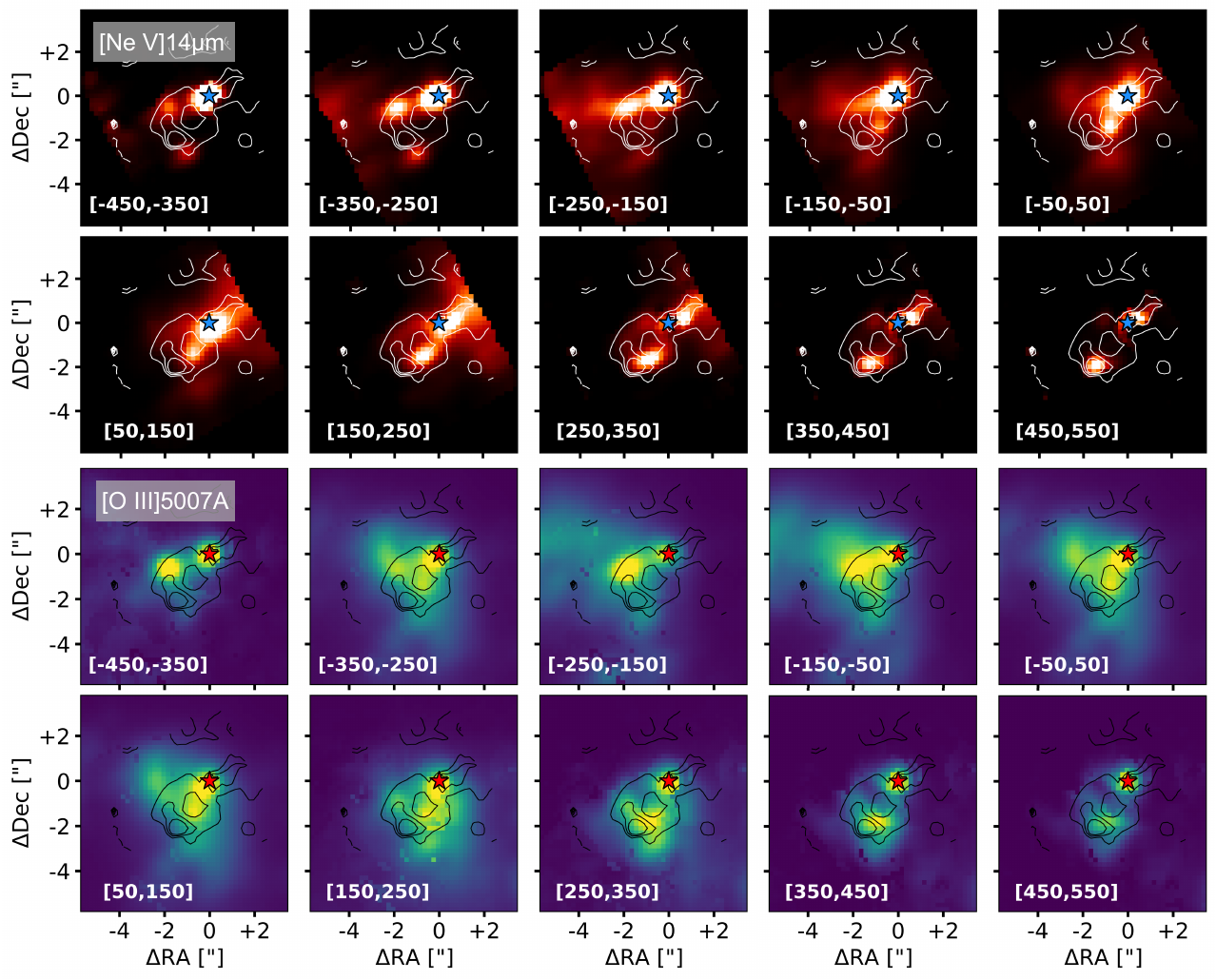}
    \caption{
    Channel maps of ionized gas species tracing the outflow kinematics from MIRI and MUSE data.
    \textit{Upper panel:} Channel maps of \nevA emission lines from MIRI data. \textit{Lower panel:} Channel maps of \oiii emission from WFM MUSE data. Contours indicate velocity dispersion levels of 97, 114, and 130 km s$^{-1}$ in the \nevA emission. Velocity bins are indicated at the top of every panel in kilometers per second and are computed relative to the same systemic velocity. The star marks the position of the nucleus based on the ALMA data (see Section \ref{app:ALMA_data}).}
    \label{fig: velocity channels}
\end{figure*}
Fig.~\ref{fig: velocity channels} presents the velocity channel maps of the \oiii and \nevA emission lines, approximately spanning the velocity range from $-$450 km s$^{-1}$ to +550 km s$^{-1}$ around the same systemic velocity. Both ionic species share the same kinematic structures in each velocity bin, although \nevA appears to trace regions closer to the outflow axis, as also found by \citet{Mingozzi2019} for high-ionization gas in MAGNUM galaxies. This is consistent with the fact that both trace the outflow but are associated with different IPs ($\sim$97~eV and $\sim$35~eV for [\ion{Ne}{V}] and [\ion{O}{III}], respectively; see Table \ref{tab: table list emission lines}). Indeed, \nevA emission arises from more internal regions of the outflow, where the AGN radiation field is expected to be stronger.
Interestingly, Fig.~\ref{fig: moment maps} shows that at projected velocities larger than 100 km~s$^{-1}$, the \nevA emission resembles the donut-shaped structure mentioned above. The bulk of the emission progressively shifts away from the nucleus with increasing channel velocity, tracing the lower arm of the donut-shaped structure.
Moreover, unlike \oiii, the redshifted \nevA velocity channels clearly reveal the receding NW ionization cone, which is otherwise highly obscured in the optical due to dust in the galaxy disk. Indeed, as is discussed in the following, the NW cone lies behind the disk and suffers stronger extinction than the SE cone \citep{Venturi2017, Venturi_2018, Marconcini2025_nat}. Thanks to the lower dust attenuation in the mid-IR, the \nevA line is an optimal tracer of this obscured outflowing component.

\subsection{Resolved diagnostic diagrams for gas excitation} \label{sec Resolved diagnostic diagrams}

\begin{figure*}[!t]
    \centering
    \includegraphics[width=1.\linewidth]{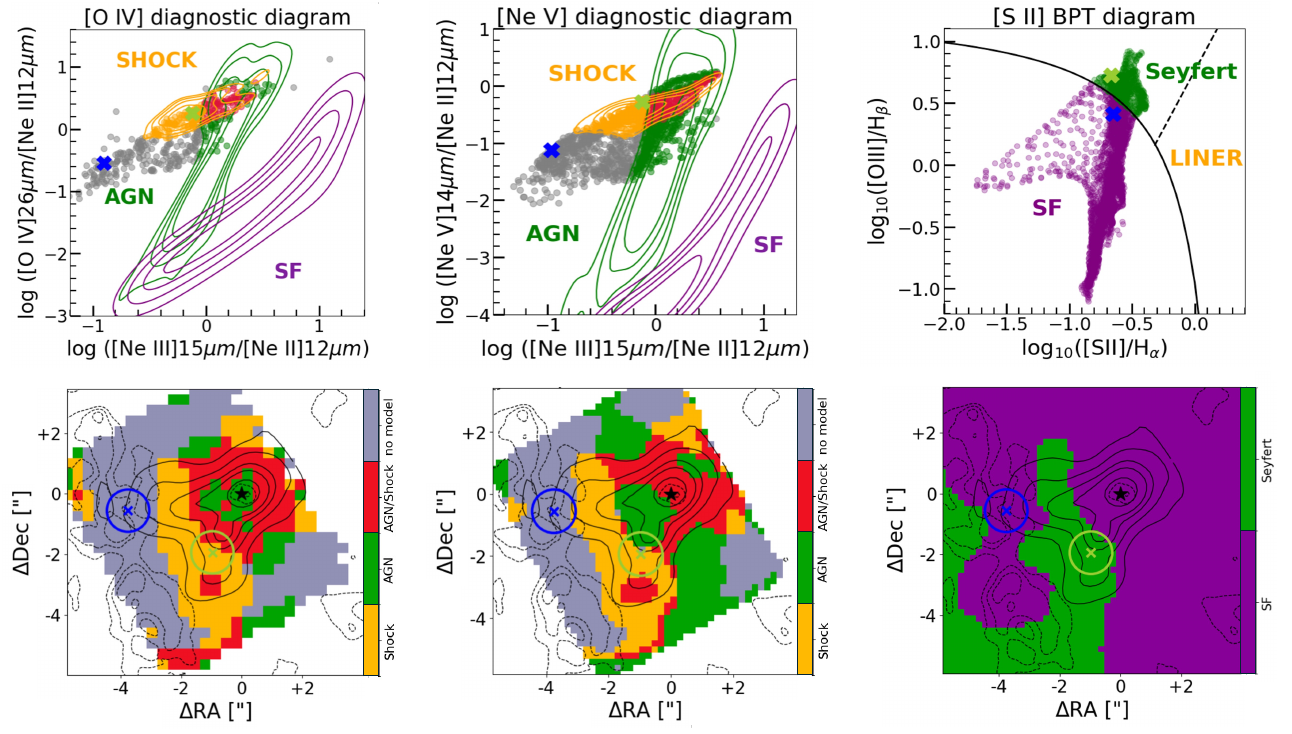}
    \caption{
    Comparison between mid-IR and optical diagnostic diagrams. Top panels, from left to right: Diagnostic diagram of [\ion{Ne}{III}]$\lambda15\mu$m/[\ion{Ne}{II}]$\lambda12\mu$m vs [\ion{O}{IV}]$\lambda14\mu$m/[\ion{Ne}{II}]$\lambda12\mu$m, [\ion{Ne}{III}]$\lambda15\mu$m/[\ion{Ne}{II}]$\lambda12\mu$m vs [\ion{Ne}{V}]$\lambda14\mu$m/[\ion{Ne}{II}]$\lambda12\mu$m, and \siiALL/H$\alpha$ vs \oiii/H$\beta$.
    In the mid-IR diagnostic diagrams, data points are color-coded by their proximity to the SF-, AGN-, and shocks-excitation models, based on the predictions by \cite{Feltre+23}. In the [\ion{S}{II}] BPT diagram, the solid curve defines the theoretical upper bound for pure SF \citep{Kewley+01}, while the dashed one separates Seyfert galaxies from LINERs \citep{Kewley+06}.
    The lower panels show spatially resolved excitation maps, where each pixel is color-coded based on its position in the corresponding diagnostic diagram.
    In the mid-IR diagnostic diagrams, shock-excited spaxels are in orange, AGN-excited spaxels in green, overlapping spaxels are in red, and spaxels not reproducible with any single model are in gray. In the [\ion{S}{II}] BPT diagram, Seyfert-excited spaxels are in green, and SF-excited spaxels are in purple.
    Solid black contours represent arbitrary \nevA flux levels, while dashed black contours represent arbitrary levels of CO(3-2) flux from ALMA data. 
    Circles are the regions where colored crosses in the upper panels and spectra in Fig.~\ref{fig: spectra} are extracted from, integrated with a radius of 0.7\arcsec.
    The star marks the position of the nucleus based on the ALMA data (see Section \ref{app:ALMA_data}). Spaxels with S/N < 10 are masked.}
    \label{fig: MIR diagnostic diagram}
\end{figure*}

Diagnostic diagrams featuring mid-IR emission line ratios are widely used to investigate the nature of the gas ionizing sources \citep[e.g.,][]{Hao2009, Weaver2010, Inami2013, Richardson2022, MartinezParedes2023, Feltre+23, Garofali2024, Mingozzi2025}.
In particular, to investigate the gas excitation in the circumnuclear region of \ngc, we employed mid-IR line ratios recently proposed by \cite{Feltre+23}, who analyzed integrated nuclear spectra of 42 local Seyfert galaxies observed with Spitzer, including \ngc. 
To ensure consistent comparisons between line ratios, we corrected the flux maps of each emission line for differences in FoV, pixel scale, and point spread functions (PSFs). In practice, for each diagnostic diagram we adopted the FoV, pixel size, and PSF corresponding to the worst case among the lines involved, as is described below.
Additionally, as is detailed in Appendix~\ref{app. MIRI corrections}, we applied multiplicative correction factors to account for flux discontinuities between the datacubes (see Fig.~\ref{fig: app stitching}).

In the upper left panel of Fig.~\ref{fig: MIR diagnostic diagram}, we show the contours of line ratio model predictions for AGN \citep{Feltre+23}, SF galaxies \citep{Gutkin2016}, and shocks \citep{Alarie2019} for the [\ion{O}{IV}]$\lambda26\mu$m/[\ion{Ne}{II}]$\lambda12\mu$m and [\ion{Ne}{III}]$\lambda15\mu$m/[\ion{Ne}{II}]$\lambda12\mu$m line ratios. 
For this diagnostic diagram, we adopted the smallest FoV (6.6\arcsec~$\times$~7.7\arcsec, i.e., Ch3), the largest pixel size (0.35\arcsec from Ch4), and convolved all images to the worst PSF (i.e., $\sim$~1.0\arcsec, the one at the reddest wavelength of $\sim26\,\mu$m). 
We include all available model grids, applying only a metallicity cut of $Z > 0.006$ (see \citealt{Feltre+23} for further details on the models).
We overlay our data on model predictions, where each pixel is color-coded based on its position in the corresponding diagnostic diagram. This diagram shows that the main ionization source in the circumnuclear region of \ngc can either be shocks or AGN, while no evidence for SF excitation is detected. According to Fig.~5 of \cite{Feltre+23}, adding fractional contributions from SF and shock models to an AGN model will shift the points toward the bottom left in the diagram. Therefore, we classify as shock- or AGN-excited only those spaxels that fall entirely within a single predicted region. Data points located in overlapping or unclassified regions cannot be uniquely attributed to a single excitation mechanism. 
To better visualize the spatial distribution and morphology of these regions, we produced diagnostic maps using the same color-coding of the diagrams, as is shown in the lower panels of Fig.~\ref{fig: MIR diagnostic diagram}. 

In the central column of Fig. \ref{fig: MIR diagnostic diagram}, we show the [\ion{Ne}{III}]$\lambda15\mu$m/[\ion{Ne}{II}]$\lambda12\mu$m versus [\ion{Ne}{V}]$\lambda14\mu$m/[Ne II]$\lambda12\mu$m diagnostic diagram. 
In this case, all lines fall within the same channel. Therefore, only the PSF convolution was required, while the FoV and pixel scale were already consistent across the maps.
Similarly to the [\ion{O}{IV}] diagram, the data points are separated into shock- and/or AGN-excited models.
The [\ion{O}{IV}] and [\ion{Ne}{V}] excitation maps reveal similar spatial structures: an extended region dominated by AGN and combined shock/AGN excitation, aligned with the disk major axis and delimited on either side by shock-dominated zones. Notably, the eastern shock front coincides with the edge of the circumnuclear ring traced by CO and H$\alpha$ (see Fig. \ref{fig: moment maps} and \citealp{Venturi_2018}), which may indicate that the outflow, emerging from the disk, interacts with the material of the circumnuclear ring, producing shocks along its path. In contrast, the centrally elongated feature may reflect direct AGN ionization within the disk.

However, \cite{Laor1998} showed that photoionizing shocks are extremely inefficient in powering the narrow line emission in AGNs. Furthermore, as has been noted by \cite{Feltre+23}, the predictions from shock models largely overlap with both AGN and star-formation grids in diagnostic diagrams. For these reasons, in Section \ref{sec: homerun} we consider only AGN and SF grids to model the optical and mid-IR emission lines.

Exploiting optical MUSE data, we compare the mid-IR diagnostic diagrams with the optical [\ion{S}{II}] BPT diagram \citep{Kewley+06,Kewley2013} shown in the right column of Fig.~\ref{fig: MIR diagnostic diagram}. Moreover, we also considered the [\ion{N}{II}] and [\ion{O}{I}] BPT diagrams, finding similar results (see Fig. 5 in \citealp{Venturi_2018} and Appendix~D3 in \citealp{Mingozzi2019} for these diagrams at large scale).
As found by \cite{Venturi_2018}, we notice that the circumnuclear SF ring seen in H$\alpha$ and in CO (see Fig.~\ref{fig: comparison fov}) is not fully dominated by SF in the BPT diagrams. Instead, as visible from the [\ion{S}{II}] spatial distribution, certain regions in the SE of the nucleus show signatures of AGN excitation. \cite{Venturi_2018} conclude that this is likely due to the superposition along the line of sight of the AGN-ionized cone and the SF ring. 
The main difference between the optical BPT and the mid-IR diagnostic diagrams lies in their sensitivity to trace the true ionization source in dusty environments. In the nuclear region, the optical BPT appears to be dominated by SF due to strong dust obscuration, as is indicated by the Balmer decrement (see Appendix~\ref{app extinction&density from muse} and Fig. \ref{fig: app extinction&density from muse}). In contrast, mid-IR diagnostics can penetrate the dust screen and clearly reveal AGN ionization at the nucleus. This explains why, in the optical, AGN ionization cones are only visible on larger spatial scales, as previously shown in \cite{Venturi_2018} and \cite{Mingozzi2019}.
Another notable difference is that the region classified as shock-dominated in the mid-IR diagrams appears to be excited by the AGN radiation according to the optical diagnostic diagram. This discrepancy likely reflects the different depths probed by optical and mid-IR emission, with mid-IR lines tracing higher ionization regions deeper into the disk, where shocks from the interaction between the outflow and the disk are expected to occur.

\begin{figure}
    \centering
    \includegraphics[width=0.9\linewidth]{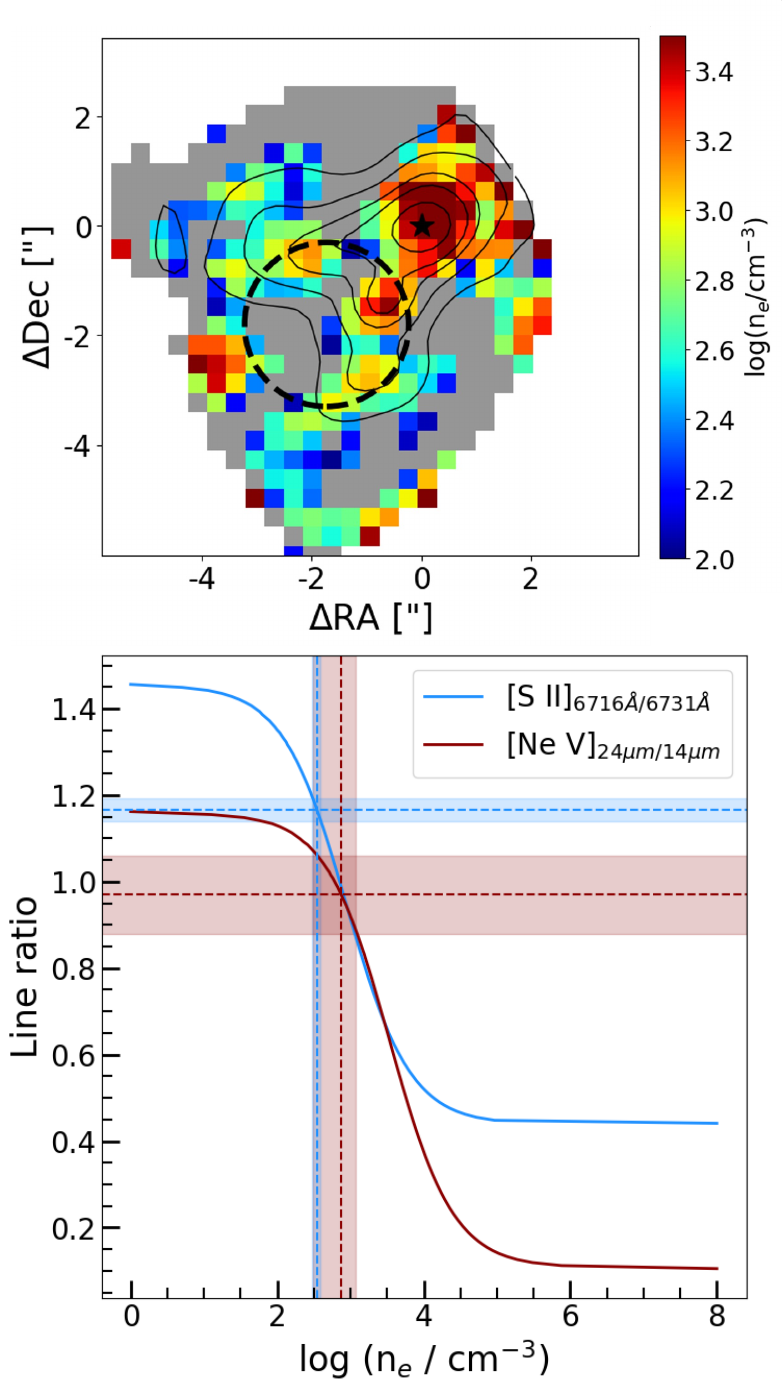}
    \caption{
    Density map from [\ion{Ne}{V}] and comparison with [\ion{S}{II}]-based values.
    \textit{Top panel}: Electron density map derived from the \nevB/\nevA\ line ratio.  Gray pixels indicate regions where the observed ratio exceeds the theoretical low-density limit (LDL) of $\sim$1.2. In these spaxels, we can assume an upper limit for the density of log(n$_e$/cm$^{-3}$)$~\le~$2. Black contours trace arbitrary flux levels of \nevA emission. The black circle marks the 1.5\arcsec radius aperture used to extract integrated densities from [\ion{S}{II}] and [\ion{Ne}{V}]. Spaxels with S/N < 10 are masked out.
    \textit{Bottom panel}: Theoretical [\ion{S}{II}] and [\ion{Ne}{V}] line ratios as a function of electron density, computed with PyNeb \citep{Luridiana+2015}. Vertical dashed lines indicate the density values in the circular aperture inferred from the observed ratios (shown as horizontal dashed lines) and shaded regions represent the errors.
    }
    \label{fig: density}
\end{figure}

\subsection{Electron density estimation} \label{sec density}
One of the main advantages of MIRI is the availability of mid-IR density-sensitive line ratios, such as \nevB/\nevA. These ratios are sensitive to a higher density regime with respect to the commonly used optical \siia/\siib ratio, due to the different ionization energies and critical densities of the lines involved. Recent studies used IR tracers to explore the gas density in local Seyfert galaxies and found densities in the range log(n$_e$/cm$^{-3}$)~=~3--5 \citep{HermosaMunoz+2024_6240,Hermosa2024, Zhang2024, Ceci_2025,RamosAlmeida+2025}.

In the top panel of Fig.~\ref{fig: density}, we show the electron density map derived from the \nevB/\nevA\ ratio using PyNeb \citep{Luridiana+2015}, assuming an electron temperature $T_e = 10^4$~K (following typical values adopted for ionized gas in AGN narrow-line regions; \citealp{Osterbrock2006}). Across the FoV, we estimate a median electron density of log($n_e$/cm$^{-3}$) = (2.7 $\pm$ 0.5), with a peak of log($n_e$/cm$^{-3}$)$\sim$3.5 at the nucleus. The density cannot be computed in some spaxels where the [\ion{Ne}{V}] line ratio exceeds the lower density limit (LDL) of $\sim$1.2. In these spaxels, we can assume an upper limit for the density of log(n$_e$/cm$^{-3}$)$~\le~$2.
In the remaining regions, the morphology of the density map appears to follow the conical structure of the outflow, as traced by the \nevA emission line (see Fig. \ref{fig: moment maps}).

Additionally, as is shown in Appendix~\ref{app extinction&density from muse}, we computed the electron density map from the \siiALL\ doublet observed in the MUSE data (see \citealp{Venturi_2018} and \citealp{Mingozzi2019} for the maps in the entire MUSE FoV). Both the circumnuclear ring and the outflow region are characterized by peaks in electron density, with values reaching log($n_e$/cm$^{-3}$)$\sim$2.6–2.8. This spatial distribution is consistent with the findings of \cite{Venturi_2018} and \cite{Kakkad2018}, who reported enhanced electron densities in similar structures.
For a quantitative comparison between the optical and mid-IR tracers, we measured the electron density from integrated spectra extracted within a 1.5\arcsec ($\sim$ 140 pc) radius aperture centered on the outflow region (see Fig. \ref{fig: density}), using both the mid-IR and optical tracers. From the mid-IR [\ion{Ne}{V}] line ratio, we estimate an electron density of (750$\pm$440)~cm$^{-3}$, while from the [\ion{S}{II}] lines we find a consistent value of (350$\pm$40)~cm$^{-3}$. These values are illustrated in the bottom panel of Fig.~\ref{fig: density}, where we plot the theoretical [\ion{S}{II}] and [\ion{Ne}{V}] line ratios as functions of electron density using PyNeb \citep{Luridiana+2015}, assuming an electron temperature $T_e = 10^4$~K.

Based on our analysis, the gas density derived from the mid-IR diagnostics is $\approx$0.3 dex higher than the one derived from the optical, although consistent within the error. This difference is in agreement with the findings of \citet{RamosAlmeida+2025}, who reported similar, or even larger, offsets in a sample of five Type-2 quasars. Following \citet{RamosAlmeida+2025}, we also attribute this discrepancy to the different physical conditions probed by the two tracers: optical lines such as \siiALL\ are more sensitive to diffuse, lower-density gas, while mid-IR fine-structure lines trace denser and more obscured regions, typically located closer to the AGN and less affected by dust extinction.

Our attempt to compute $n_e$ from the \arvB/\arvA ratio within the 1.5\arcsec-radius aperture interestingly yielded a ratio of $\sim$2.1, which is above the [\ion{Ar}{V}] LDL value of $\sim$1.8 (corresponding to an upper limit on log(n$_e$/cm$^{-3}$) density similar to the [\ion{Ne}{V}] one, $\sim$~2).
More importantly, even when integrating the [\ion{Ne}{V}] flux across the entire MIRI FoV, we do not observe line ratios falling in the LDL, contrary to what was found by \citet{Dudik+07} using Spitzer/IRS data. This further supports the idea that their higher \nevB/\nevA ratios - and apparent detection of LDL conditions - may originate from observational limitations.
In particular, the higher sensitivity of JWST/MIRI MRS compared to {\it Spitzer}/IRS may have allowed for a more effective detection of low surface brightness emission that could have remained undetected in previous studies.
A direct comparison with the \citet{Dudik+07} measurements and observational setup is presented in Appendix~\ref{app dudik LDL}.

\subsection{Photoionization modeling with HOMERUN} \label{sec: homerun}

We modeled the full set of optical to mid-IR emission using state-of-the-art photoionization models. Specifically, we employed the Highly Optimized Multi-cloud Emission-line Ratios Using photo-ionizatioN (HOMERUN) modeling framework developed by \cite{Marconi+2024}. 

\subsubsection{Model grids of AGNs and SF}
Given the superposition along the line of sight of the AGN-ionized cone and the circumnuclear SF ring, in the HOMERUN fit we combined two suites of models: one reproducing the emission of dust-free gas ionized by the AGN, and another accounting for the SF contribution, modeled as emission from dusty nebulae around \ion{H}{ii} regions and including both dust depletion and dust physics. Both components were assumed to have the same metallicity; however, while we adopted dust-free models for the AGN component, we included dust in the SF models and corrected gas-phase metallicities using the depletion factors listed in Table 7.8 of Hazy 1 (Cloudy v23.1). This choice was motivated by the optical extinction map (Appendix~\ref{app extinction&density from muse}), which shows clear dust structures associated with the SF ring, but not with the outflowing AGN-ionized gas. Furthermore, the AGN-ionized component exhibits iron coronal lines, indicating that at least part of the dust has been destroyed, releasing iron into the gas phase.

Specifically, the shape of the AGN ionizing radiation field is described by a power law with UV slope $\alpha_{UV}=-0.5$, an exponential cutoff exp(-h$\nu$/k T$_{\rm Max}$) and the X-ray component slope of $\alpha_{X}$=-1.0 linked to the UV through the $\alpha_{ox}$. We have computed models with the following combinations of log(T$_{\rm Max}$/K) = 4.0, 4.5, 5.0, 5.5, 6.0, 6.5, 7.0 and $\alpha_{ox}=-1.2, -1.5, -1.8$. 

As is described in \cite{Marconi+2024}, the incident spectra for the SF model grids are stellar population models from BPASS v2.3 \citep{Byrne2022, Stanway&Eldridge2018} including binary stellar evolution. These models use a \cite{kroupa2001} initial mass function and an upper mass cutoff of 300 M$_{\odot}$. We selected models with solar value for the stellar metallicity log(Z$_{star}$)=-1.7 and ages in log(age/Myr) of 6.0, 6.4, 6.6, 7.1, 8.8.

The other model parameters are identical for the two suites of models (AGN and SF); namely, the ionization parameter log(U) (i.e., the number of ionizing photons compared to that of atoms of neutral hydrogen) ranges from $-4.0$ to $-0.5$ in step of 0.5 and the hydrogen density of the gas, log(n$_{\rm H}$~/~cm$^{-3}$) from 0 to 7 in steps of 1. 
For the first iteration HOMERUN spans metallicities log(Z/Z$_{\odot}$) from -1.0 to 0.4 in steps of 0.2. After this first iteration, HOMERUN performs a final refinement in steps 0.02 on predictions interpolated across the finer grid.

\subsubsection{HOMERUN modeling approach}
The HOMERUN models the line emission with a weighted combination of multiple CLOUDY single-cloud photoionization models \citep{Ferland1998}, iterating over different gas metallicities and incident radiation fields.
The novelty of this method is that the weights assigned to the individual single-cloud models are treated as free parameters during the fitting process.

As is explained in detail in \cite{Marconi+2024}, the fitting process starts by selecting a grid of CLOUDY models, defined by hydrogen density ($n_\mathrm{H}$) and ionization parameter ($U$) for a fixed ionizing spectrum ($S_\nu$), gas phase oxygen abundance ($A_\mathrm{O}$), and elemental abundance ratios ($Z$). For each grid, the code searches for the best-fitting multicloud model by combining individual single-cloud templates.
This is achieved by solving a non-negative least squares problem, whereby the weights of each cloud are constrained to be non-negative. The procedure is conceptually analogous to stellar population fitting methods (e.g., pPXF; \citealt{Cappellari2004}), where a galaxy spectrum is reproduced as a linear combination of stellar templates with positive weights. 
No regularization is applied in the minimization, allowing for full flexibility in the weight distribution. The process is repeated across all available grids, spanning different ionizing continua, metallicities, and abundance scalings. 
The optimal solution is identified as the one minimizing the loss function $L_\mathrm{min}(S_\nu, A_\mathrm{O}, Z)$, which quantifies the deviation between model predictions and observed emission-line fluxes. This loss function is effectively a reduced $\chi^2$ statistic, so values $\lesssim 1$ indicate a statistically good fit.

When exploring different metallicities, elemental abundances were scaled from the solar photospheric values of \citet{Asplund2021}, except for carbon and nitrogen, which were rescaled following the prescriptions of \citet{Nicholls2017}. During the fit, all elemental abundances except oxygen were allowed to vary from their initial values. 
Scaling factors were used for the AGN and SF components. Since the AGN component was assumed to be dust-free, while the SF component includes dust, refractory elements can have different scaling factors (due to depletion), whereas non-refractory elements share the same scaling factor in both components.

\subsubsection{HOMERUN fitting results} \label{sec HOMERUN fitting results}

\begin{figure}
    \centering
    \includegraphics[width=1\linewidth]{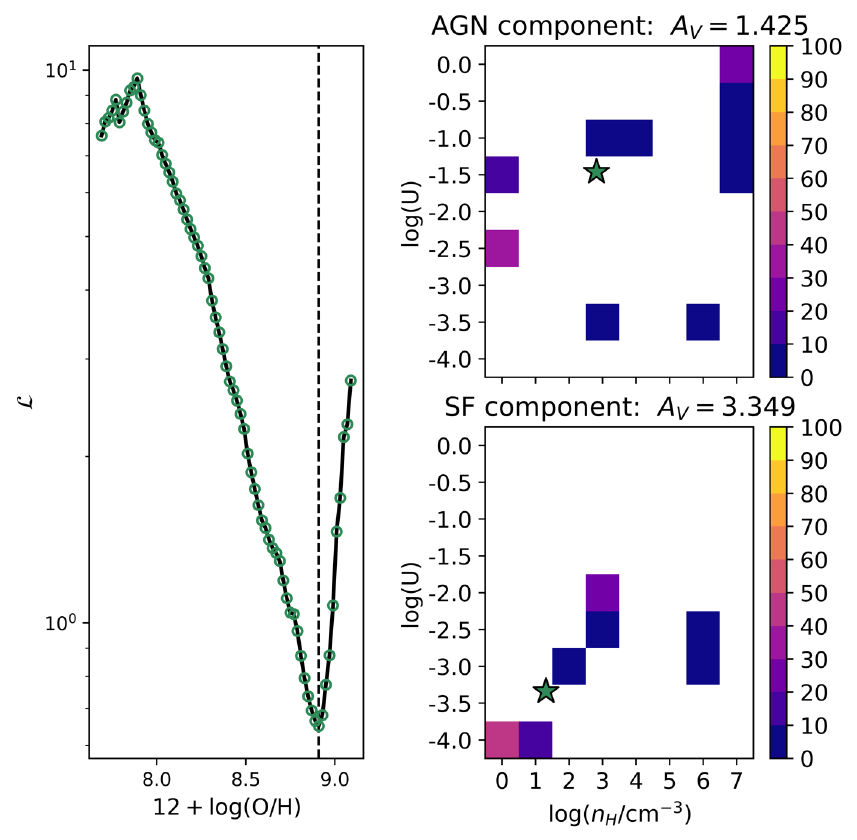}
    \caption{HOMERUN model results using two component.
    \textit{Left panel: }Variation in the loss function, $\mathcal{L}$, as a function of the oxygen abundance, 12+log(O/H). In this scale, the solar metallicity is 8.69. The vertical dashed black line represents the metallicity of the model with the minimum value of $\mathcal{L}$.
    \textit{Right panels: }Grids of single-cloud models in log(U) and log(n$_H$) for the two components of the fit. The colors represent the weights of each single-cloud model, as indicated by the color bar, when $\mathcal{L}$ reaches the minimum value. The star represents the weighted density and ionization parameter of the single-cloud models.
    }
    \label{fig: homerun results}
\end{figure}

We fit the fluxes of 60 ionized and neutral atomic emission lines measured from the MIRI and MUSE spectra (32 from MIRI and 28 from MUSE, see Figs. \ref{fig: spectra} and \ref{fig: app MUSE spectra}) extracted within a 1.5\arcsec radius aperture centered on the outflow region (see Fig. \ref{fig: density}). 
This aperture was chosen as the largest circular region that is fully covered by all MIRI channels, ensuring the detection of the maximum number of emission lines.
The unique approach of HOMERUN enables us to reproduce at the same time 60 different emission lines in different ionization stages, with the majority of them reproduced with better than 15$\%$ accuracy.
The data are best matched by a SF component with mean $\rm A_V$ = 3.3, $\rm log(n_H$~/~cm$^{-3}$) = 1.3, and $\rm log(U)$ = -3.3 and an AGN component with mean $\rm A_V$ = 1.4,  $\rm log(n_H$~/~cm$^{-3}$) = 2.9, and $\rm log(U)$ = -1.4 (see Fig.~\ref{fig: homerun results}). Being results of ionized gas region, in the following we will consider $n_H = n_e$.

We find that the SF component is more dust-attenuated than the outflow, in agreement with the extinction map obtained from the Balmer decrement (see Fig. \ref{fig: app extinction&density from muse}) where the higher A$_V$ values trace the circumnuclear ring and the nucleus. Interestingly, the difference in A$_V$ between SF and AGN components is in line with the median values found in MAGNUM galaxies dividing the emission in disk and outflow \citep{Mingozzi2019}.
In Section \ref{sec density}, we computed the electron density from the mid-IR [\ion{Ne}{V}] and the optical [\ion{S}{II}] line doublet ratios within the same 1.5\arcsec radius aperture. The density derived from [\ion{Ne}{V}], $\rm log(n_e$/cm$^{-3}$) $\sim$ 2.87, is consistent with the average value of the AGN model components, $\rm log(n_e$/cm$^{-3}$) = 2.9. This is expected since HOMERUN attributes the entire emission of [\ion{Ne}{V}] to the AGN (see Table \ref{tab:homerun emission_lines 2}), as stars may not produce hard-enough radiation to account for the emission of this transition. 
Conversely, the [\ion{S}{II}] lines yield a higher density of log(n$_e$/cm$^{-3}$)~$\sim$~2.5 within the 1.5\arcsec radius aperture, compared to log(n$_e$/cm$^{-3}$)~$\sim$~1.3 from HOMERUN. 
This is because the average density of the SF component from HOMERUN is lower than that from the [\ion{S}{II}] doublet and  HOMERUN requires a contribution of more than 70$\%$ from the SF component to reproduce the [\ion{S}{II}] emission (see Table \ref{tab:homerun emission_lines 1}).
This highlights the strength of our decomposition method: while classical [\ion{S}{II}] diagnostics measure a global electron density, HOMERUN disentangles the different ionizing sources and shows that the SF component alone has a much lower density.
Finally, the derived ionization parameters are log(U)~=~−1.4 for the AGN component and a lower log(U) of −3.3 for the SF component, both in line with values found in H~II regions (e.g., \citealp{Kewley+01, Dopita2006}) and AGN narrow-line regions (e.g., \citealp{Groves2004}). This supports the effectiveness of our decomposition in isolating distinct ionization regimes.

Figure~\ref{fig: homerun results} shows the results of the fit. The left panel shows the behavior of the loss function as a function of oxygen abundance. The minimum of the curve is found for the minimum loss function of $\mathcal{L}_\mathrm{min}$ = 0.6, corresponding to 12+log(O/H)~=~8.91.
The left panels represent the weights of the single-cloud models with different U and N$_H$ which were derived for the best fit, for the AGN and SF components in the upper and lower panel, respectively. Using the weights, it is possible to compute the average density and ionization parameter of the clouds, which are identified by the green star.
Only a small number of single-cloud models have non-zero weights, and the AGN and SF components contribute differently across the various lines, depending on the ionization state. 
In particular, as is indicated in Tables~\ref{tab:homerun emission_lines 1}~and~\ref{tab:homerun emission_lines 2}, high-ionization lines (e.g., \ion{Ne}{V}, \ion{O}{III}) are predominantly reproduced by the AGN component, while lower-ionization lines (e.g., \ion{N}{II}, \ion{Cl}{II}, \ion{Ar}{II}) are mainly explained by the SF component.
Interestingly, the H$\alpha$ luminosity is produced more than 70$\%$ by AGN excitation.

Using the same procedure, we fit with HOMERUN the emission lines in the spectra extracted from the regions marked in Fig.~\ref{fig: comparison fov} (see Figs.~\ref{fig: spectra} and \ref{fig: app MUSE spectra}). In the outflow-dominated region (green spectrum), we find for the SF component $\rm A_V$ = 2.8, $\rm log(n_H$~/~cm$^{-3}$) = 0.9, and $\rm log(U)$ = -3.4, and for the AGN component $\rm A_V$ = 1.4, $\rm log(n_H$~/~cm$^{-3}$) = 3.3, and $\rm log(U)$ = -1.4. In the stellar disk region (blue spectrum), we find for the SF component $\rm A_V$ = 2.9, $\rm log(n_H$~/~cm$^{-3}$) = 1.4, and $\rm log(U)$ = -3.3, and for the AGN component $\rm A_V$ = 1.5, $\rm log(n_H$~/~cm$^{-3}$) = 5.5, and $\rm log(U)$ = -1.5. These values are consistent with those obtained from the spectrum extracted within a 1.5\arcsec\ radius aperture (see Fig.~\ref{fig: density}). Moreover, SF emission accounts for 64$\%$ of the H$\beta$ luminosity in the outflow region and 88$\%$ in the stellar disk region.

\subsubsection{Ionized gas mass from HOMERUN} \label{sec Ionized gas mass from HOMERUN}
A key outcome of the HOMERUN fitting procedure is the ability to relate the luminosity of an emission line to the mass of the ionized gas associated. Since the fit is based on a physically consistent combination of AGN and SF single-cloud models, it allows us to disentangle the contributions from the different ionizing sources and connect the observed emission to intrinsic gas properties. In this way, under the assumption that the outflowing gas is fully ionized by the AGN, we avoided including the contribution from systemic or SF-related components, ensuring that the derived masses trace the outflowing gas itself. 
In particular, for each component, HOMERUN provides both the intrinsic (i.e., extinction-corrected) luminosities (see Tables~\ref{tab:homerun emission_lines 1}~and~\ref{tab:homerun emission_lines 2}) and the corresponding ionized gas mass. From the underlying single-cloud \textsc{CLOUDY} models, one obtains for each model a surface brightness in H$\beta$ (i.e., $\log L_{\mathrm{H}\beta}/\mathrm{area}$) and a gas mass surface density (i.e., $\log M_{\mathrm{ion}}/\mathrm{area}$). These model outputs define a direct proportionality between H$\beta$ luminosity and ionized gas mass.
By rescaling this relation to the total luminosity of a specific emission line (e.g., [\ion{Ne}{V}] or [\ion{O}{III}]) and taking into account the AGN fraction, $f_{\mathrm{AGN}}$, as determined from the HOMERUN fit, we derive a general expression:
\begin{equation}
    M_\mathrm{ion} = C_{\mathrm{line}} \times \frac{L_{\mathrm{line}}}{f_\mathrm{AGN}},
\end{equation}
where $C_{\mathrm{line}}$ is a calibration coefficient derived from the HOMERUN AGN component models.
This framework provides a robust and physically grounded method to estimate the mass of ionized outflowing gas from spatially integrated line luminosities. 

After applying HOMERUN to \ngc emission line fluxes, we found the following ionized gas masses:
\begin{align}
    M_{[O\,III]} &= 0.74\times 10^5 M_\odot \left(\frac{L_{AGN}([O~III])}{10^{40} erg\,s^{-1}}\right) \label{homerun mass oiii}\\
     &= 3.0 \times 10^5 M_\odot \left(\frac{L_{obs}([O~III])}{10^{40} erg\,s^{-1}}\right), \label{homerun mass oiii 2}
\end{align}
\begin{align}
    M_{[Ne V]} &= 4.2 \times 10^5 M_\odot \left(\frac{L_{AGN}([Ne~V])}{10^{40} \text{ erg s}^{-1}}\right) \label{homerun mass nev}\\
               &= 4.4 \times 10^5 M_\odot \left(\frac{L_{obs}([Ne~V])}{10^{40} \text{ erg s}^{-1}}\right), \label{homerun mass nev 2}
\end{align}
where $L_{AGN}([O~III])$ and $L_{AGN}([Ne~V])$ are the intrinsic line luminosities emitted by the AGN component, and $L_{obs}([O~III])$ and $L_{obs}([Ne~V])$ are the total observed luminosities (i.e., without the HOMERUN separation between the AGN and SF). The scaling factors applied to the observed luminosities account for both the fractional contribution of the AGN component and the effect of reddening. The scaling factors of $L_{AGN}$ in Equations~\ref{homerun mass oiii}~and~\ref{homerun mass nev} refer to extinction-corrected luminosities and can be directly compared with the standard conversion factors adopted in analytical calculations of the ionized gas masses (Equations~\ref{eq: M outflow NeV}~and~\ref{eq: M outflow OIII}, see also the discussion in the next section).

We computed an outflow mass of  $\rm M_{[Ne V]}$~= (109~$\pm$~2)~$\times$~10$^{3}$~\msun and $\rm M_{[O III]}$~=~(88~$\pm$~9)~$\times$~10$^{3}$~\msun within the 1.5\arcsec radius aperture. 
These differences in mass, obtained with HOMERUN, reflect the discrepancy between the observed and model-predicted line ratios. Indeed, the observed ratio between \oiii and \nevA fluxes is 0.81 times the model ratio (see Appendix~\ref{app homerun results}), which matches the ratio between the \oiii- and \nevA-based outflow masses.
We emphasize that the scaling relations in Equations~\ref{homerun mass oiii}-\ref{homerun mass nev 2} for the ionized outflow masses are not universally applicable to other AGNs. They are derived specifically from the HOMERUN multicloud modeling customized for the physical conditions and emission properties of NGC 1365. Therefore, applying these relations to other galaxies requires performing a dedicated HOMERUN fit tailored to the individual characteristics of each galaxy. In the next section, we compare the outflow mass obtained with HOMERUN modeling with that estimated following a standard and commonly used approach (e.g., \citealt{CanoDiaz2012, Carniani2015}, see Appendix~\ref{app mass estimation} and Table \ref{tab:outflow_properties}).

\subsection{Outflow energetics} \label{sec Outflow energetics}

\begin{figure*}[!]
    \centering
    \includegraphics[width=.7\linewidth]{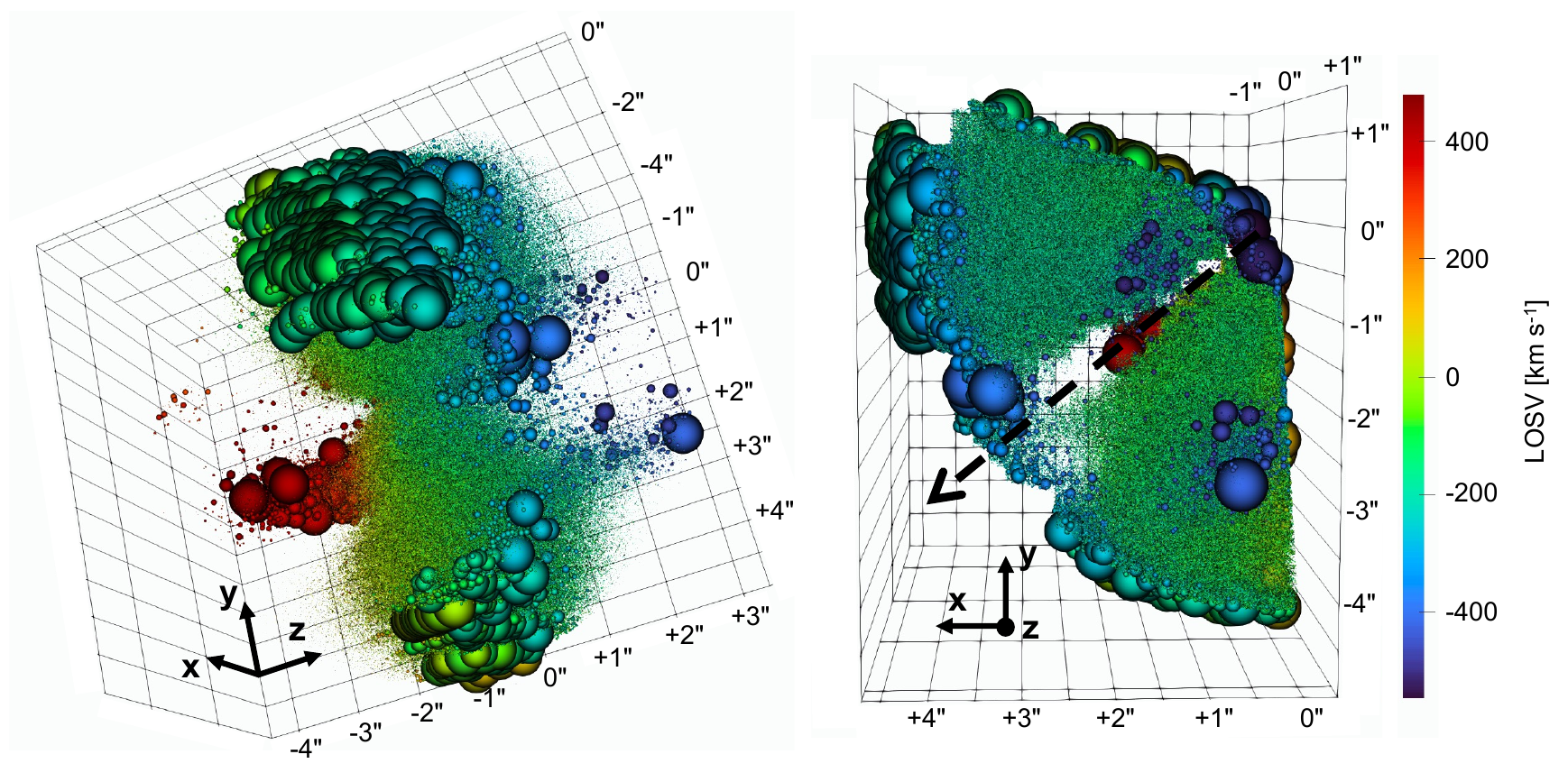}
    \caption{
    Three-dimensional reconstruction of the \MOKA\ best-fit model of the ionized outflow traced by \nevA. Gas clouds are color-coded by their LOSV. The two panels show different viewing angles: the left one provides a view into the inner cone region, while the right one corresponds to the observer's line of sight. The XY plane represents the plane of the sky, with the Y and X axes oriented in the northern and eastern directions, respectively, while the Z axis corresponds to the line of sight. In the right panel, the dashed arrow indicate the direction of the conical outflow axis. According to the color bar, blue and red clouds are blueshifted and redshifted, respectively. Bubble size scales with the intrinsic flux of each cloud. 
    }
    \label{fig: moka model}
\end{figure*}

In the previous section, with HOMERUN we obtain an outflow mass of $\rm M_{[Ne V]}$~=~(109~$\pm$~2)~$\times$~10$^{3}$~\msun and $\rm M_{[O III]}$~=~(88~$\pm$~9)~$\times$~10$^{3}$~\msun within a 1.5\arcsec radius aperture centered on the outflow region (see Fig. \ref{fig: density} and Section \ref{sec HOMERUN fitting results}). 
In addition, following \cite{Carniani2015}, in Appendix~\ref{app mass estimation} we present a detailed derivation of the outflow mass of \nevA and \oiii. Overall, with the classic method we find in the same spatial region $\rm M_{[Ne V]}$~=~(2.6~$\pm$~1.0)~$\times$~10$^{3}$~\msun and $\rm M_{[O III]}$~=~(20~$\pm$~7)~$\times$~10$^{4}$~\msun.
The large difference between the HOMERUN and classical mass estimates (factors of $\sim$40 for [\ion{Ne}{V}] and $\sim$2 for [\ion{O}{III}]) clearly highlights the fundamental limitations of the standard approach. The classical method relies on several strong assumptions, such as considering that all the species are in the ionization state responsible for the observed line (e.g., O$^{2+}$ for [\ion{O}{III}], Ne$^{4+}$ for [\ion{Ne}{V}]), adopting a single average density (from [\ion{S}{II}] in the optical and [\ion{Ne}{V}] in the mid-IR), and correcting for attenuation using the Balmer decrement. 
In contrast, HOMERUN does not require assumptions about ionization levels, temperature, or metallicity, and accounts for differential reddening between the two components. 
The reliability of the more complex HOMERUN approach is demonstrated by the fact that classical method mass estimates differ by a factor of $\sim$75 between different tracers, highlighting the inherent inconsistencies of the traditional approach.
Indeed, for example, the HOMERUN fitting shows that the fraction of neon in the form of Ne$^{4+}$ in the best-fit models is only $\sim$10$\%$, while for O$^{2+}$ the fraction is significantly higher, around 60$\%$ of the total oxygen. This is certainly one of the main factors contributing to the large discrepancy between the two mass estimates. 
Moreover, it should be noted that [\ion{O}{III}] and [\ion{Ne}{V}], having different critical densities and luminosities, trace gas from slightly different physical regions within the outflowing clouds, which further contributes to the observed differences in mass estimates beyond the methodological inconsistencies of the classical approach.

We assume that the flux density within each spaxel is constant over time and that the outflow subtends a solid angle, $\Omega$. Therefore, we computed the mass outflow rate as $\dot{M}_{\rm out} = M_{\rm out} v_{\rm out} / \Delta R$, i.e., the amount of ionized mass crossing a distance, $\Delta R$, with velocity, $v_{\rm out}$. We also calculated the outflow kinetic energy as $E_{\rm out} = \frac{1}{2} M_{\rm out} v_{\rm out}^2$ and the outflow power as $P_{\rm out} = \frac{1}{2} \dot{M}_{\rm out} v_{\rm out}^2$.

\begin{table*}[!ht]
    \centering
    \caption{Properties of the ionized outflow in NGC 1365 traced by \nevA and \oiii\ emission lines.}
    \label{tab:outflow_properties}
    \begin{tabular}{llccccc}
        \toprule\toprule
        Line & Method                       & ${M}_{\mathrm{out}}$   &  $\dot{M}_{\mathrm{out}}$     & $E_{\mathrm{out}}$  & $\dot{E}_{\mathrm{out}}$ \\
            &                               & [10$^{4}$ M$_\odot$]   & [10$^{-1}$ M$_\odot$ yr$^{-1}$] & [10$^{53}$ erg]   & [10$^{39}$ erg s$^{-1}$] \\
        \midrule
        \multirow{2}{*}{\nevA} & standard   & 0.26 $\pm$ 0.10        & 0.041 $\pm$ 0.020              & 0.04 $\pm$ 0.02    & 0.20 $\pm$ 0.13 \\
                               & HOMERUN    & 10.9 $\pm$ 0.2         & 1.7 $\pm$ 0.2                 & 1.6 $\pm$ 0.3      & 8 $\pm$ 3 \\
        \addlinespace
        \multirow{2}{*}{\oiii} & standard   & 20 $\pm$ 7             & 3.8 $\pm$ 0.18                  & 4 $\pm$ 2       & 26 $\pm$ 17 \\
                               & HOMERUN    & 8.8 $\pm$ 0.9          & 1.7 $\pm$ 0.3                  & 1.9 $\pm$ 0.5       & 12 $\pm$ 4 \\
        \bottomrule
    \end{tabular}
    \tablefoot{From left to right: Emission line, method used to compute the energetics, outflow mass, mass outflow rate, kinetic energy, and power of the ionized outflow. These values are derived from the spatially resolved outflow properties in the FoV of MIRI Ch3, using the \MOKA\ tool (see details in Section \ref{sec Outflow energetics}).}
\end{table*}

We inferred the \nevA and \oiii intrinsic (i.e., de-projected) radial velocities of the outflow (v$_\mathrm{out}$) using the 3D multicloud \MOKA kinematic model \citep{Marconcini2023}. This model has demonstrated exceptional power in reproducing outflow features with unprecedented detail, leveraging IFU data from a variety of instruments \citep{Marconcini2023, Cresci2023,Perna2024,Ulivi2025, MIRACLE_NGC424, Marconcini2025_nat}. To reproduce the outflow features in NGC 1365, we adopted a conical outflow morphology with maximum extension of 5 arcsec ($\sim$ 470 pc) and position angle\footnote{The position angle is measured clockwise from north.} of 235$^{\circ}$. The free parameters of the fit are the inclination of the cone axis with respect to the line of sight ($\beta$) and the intrinsic outflow velocity (v$_{\rm out}$). With \MOKA \ we reproduce the line emission on a spaxel-by-spaxel basis, which as a consequence guarantee that the model reproduces the observed moment maps, with discrepancies between observed and modeled moment maps below 5$\%$.
We found that the best-fit parameters that reproduce the \nevA outflow features in \ngc are $\beta$ = 82 $\pm$ 6$^{\circ}$, and v$_{\rm out}$ = 390 $\pm$ 35 km s$^{-1}$, consistent with the results in \cite{Marconcini2025_nat}. Fig. \ref{fig: moka model} shows the 3D representation of this conical outflow.
Moreover, we used \MOKA to fit the \oiii spatially resolved emission assuming the same outflow geometry as for the \nevA emission and found $\beta$ = 84 $\pm$ 5$^{\circ}$, and v$_{\rm out}$ = 470 $\pm$ 40 km s$^{-1}$. 
These results are also valid in the region where we extracted the spectrum used for the HOMERUN fitting. 

So, adopting $\Delta$R = 284 pc ($\sim$ 3\arcsec, i.e., the sky-projected extension of the 1.5\arcsec radius aperture), we obtained the results of the outflow energetics reported in Table \ref{tab:outflow_properties}.
From HOMERUN, we calculated consistent mass outflow rates for \nevA and \oiii, namely $\dot{\rm M}_{[Ne~V]}$ = $\left(1.72~\pm~0.20\right)~\times~10^{-1}$ M$_\odot$~yr$^{-1}$ and $\dot{\rm M}_{[O~III]}$ = $\left(1.68~\pm~0.31\right)~\times~10^{-1}$ M$_\odot$~yr$^{-1}$, respectively. 
Instead, with the classic approach the [\ion{O}{III}]-derived outflow rate is two orders of magnitude larger than that from [Ne~V]. 
This discrepancy is primarily driven by the underestimation of the [\ion{Ne}{V}] outflow mass in the classical approach. Specifically, the classical [\ion{Ne}{V}] outflow mass is $\sim$75 times smaller than the [\ion{O}{III}] estimate and $\sim$40 times smaller than the HOMERUN value, while the classical [\ion{O}{III}] mass is only a factor of $\sim$2 higher than the HOMERUN mass. This asymmetric bias clearly demonstrates that the limitations of the standard method predominantly affects high-ionization tracers like [Ne~V], leading to severely underestimated masses and consequently unreliable mass outflow rate calculations.

Our mass outflow rate estimate derived with \MOKA+HOMERUN models is about one order of magnitude larger than the total ionized outflow rate reported by \cite{Venturi_2018}, which was derived from the broad H$\alpha$ component in a shell-like region of the SE cone, extending from 1\arcsec to 6\arcsec from the nucleus.
This large discrepancy clearly highlights how significantly the adopted methodology can affect the inferred outflow properties.

\section{Conclusions}\label{sec conclusion}
In this work, we have presented JWST/MIRI MRS observations of the active galaxy \ngc as part of the Mid-IR Activity of Circumnuclear Line Emission (MIRACLE) program. We analyzed the spatially resolved gas properties and compared them with optical and millimeter observations from MUSE and ALMA, respectively.
We traced both the ionized atomic and the warm and cold molecular gas phases in the circumnuclear region ($\sim$ 0.9$\times$0.9 kpc$^2$) of the \ngc galaxy, identifying more than 40 mid-IR emission lines. 
Our main results are the following:
\begin{itemize}
    \item We find that the mid-IR, optical, and millimeter emission lines consistently trace different phases of the gas with coherent internal kinematics within each phase (see Fig. \ref{fig: moment maps} and Section \ref{sec Multi-phase gas kinematics}).
    The cold molecular gas traced by the CO(3-2) transition shares the same ordered kinematics of the stellar galaxy disk \citep{Venturi_2018}, with no evidence of outflowing material. The same conclusion holds for the warm molecular gas traced by the H$_2$ pure-rotational transitions, with the exception of a peculiar high-$\sigma$ unresolved region located southward of the nucleus.
    \item The ionized gas lines can be separated in two main groups that show opposed morphologies based on their IP. The high-IP ($>$~54~eV) emission lines trace the bipolar outflow, showing a velocity gradient perpendicular to the disk major axis, consistent with the \oiii kinematics. The velocity dispersion map highlights a donut-shaped structure associated with the approaching side of the outflow, which extends up to 2.5\arcsec ($\sim$ 240 pc).
    The low-IP ($<$~25~eV) emission line kinematics resembles the stellar galaxy disk ordered motions, similar to the H$_2$, with clumpy flux peaks in the circumnuclear star-forming ring. 
    Finally, medium-IP lines show intermediate properties, with a  kinematics similar to low-IP species but exhibiting the donut-shaped structure typical of the high-IP lines in the velocity dispersion map.
    
    \item \oiii and \nevA velocity channel maps reveal that both outflowing species trace the same kinematic structures, with the \nevA appearing more collimated on the outflow axis, as was expected from their different IPs (Fig. \ref{fig: velocity channels} and Section \ref{sec Velocity channel maps}). The \nevA emission also reveals the NW receding counterpart of the ionization cone, which is undetected in \oiii due to large extinction caused by the galaxy disk at optical wavelengths.
    
    \item The mid-IR emission-line ratios indicate a composite excitation scenario, with contributions from both AGN photoionization and shocks (Fig. \ref{fig: MIR diagnostic diagram} and Section \ref{sec Resolved diagnostic diagrams}). The AGN ionization seems to extend along the disk major axis, while in the other direction we observe shock-dominated zones, suggesting an interaction between the outflow and the disk material. Notably, by penetrating the dust in the central regions of the galaxy, these mid-IR diagnostic diagrams reveal for the first time the role of the AGN in ionizing the gas in the nuclear region, differently from the optical diagnostic diagrams that mostly reveal excitation due to SF.
    
    \item The spatially resolved electron density map derived from the \nevB/\nevA line ratio shows a median density of (750~$\pm$~440)~cm$^{-3}$ (Fig. \ref{fig: density}), $\approx$0.3 dex higher than the value inferred from optical \siia/\siib line ratio (see Sections \ref{sec density} and Appendix~\ref{app extinction&density from muse}).

    \item By exploiting the innovative photoionization modeling code HOMERUN, we derived the physical properties of the ionized outflowing gas, providing a self-consistent and physically motivated estimate of the outflow mass (Section~\ref{sec: homerun}). We obtained $\rm M_{[Ne\,V]}$~=~(109~$\pm$~2)~$\times$~10$^{3}$~\msun and $\rm M_{[O\,III]}$~=~(88~$\pm$~9)~$\times$~10$^{3}$~\msun. These estimates differ significantly from those derived with the standard method, which are $\sim$40 times lower ([\ion{Ne}{V}]) and $\sim$2 times higher ([\ion{O}{III}]) than those obtained from HOMERUN. This discrepancy reflects the strong assumptions in the classical approach regarding ionization state, average density, and extinction correction.

    \item We find a consistent outflow mass rate of $\sim$0.17 M$_\odot$ yr$^{-1}$ from  \nevA and \oiii (Section \ref{sec Outflow energetics}). Our estimate is about one order of magnitude higher than the value reported in the literature, clearly illustrating how different methods can significantly impact the inferred outflow properties. To derive the intrinsic outflow velocities, we used the 3D \MOKA\ kinematic model, assuming a conical geometry. The best-fit solutions provide consistent inclination angles ($\beta$~$\sim$~82-84$^\circ$) and outflow velocities (v$_\mathrm{out}$~$\sim$~390-470~km~s$^{-1}$) for both \nevA and \oiii, reinforcing the reliability of our kinematic and energetic estimates.

\end{itemize}

This work highlights the critical need for self-consistent, multiline photoionization and kinematical modeling to properly capture the complex structure and energetics of AGN-driven outflows -- a key step toward understanding their impact on galaxy evolution.
We fully exploited the synergy between cutting-edge IFU facilities, JWST/MIRI and VLT/MUSE, to model the ionized gas phase properties across a broad ionization range (from $\sim$10 eV to $\sim$130 eV), using a total of 60 emission lines. 
For the first time, we combined detailed 3D kinematic modeling with \MOKA and photoionization fitting with HOMERUN to derive the intrinsic structure and physical conditions of the outflow in a fully self-consistent framework. This unprecedented approach enabled a coherent and physically motivated characterization of the outflow’s spatial structure, ionization stratification, and energetics.
Our results show that classical methods, based on simplified assumptions and a limited set of tracers, significantly underestimate the outflow energetics -- and hence the strength of AGN feedback.

\section{Data availability}
The moment maps of the other emission lines listed in Table \ref{tab: table list emission lines} are shown in the \href{https://zenodo.org/records/17535125}{online material}.
The reduced MIRI IFU datacubes used in this work are available at the CDS via anonymous ftp to \url{cdsarc.cds.unistra.fr (xxxxxxxx)} or via \url{https://cdsarc.cds.unistra.fr/viz-bin/cat/J/A+A/xxx/xxx}.

\begin{acknowledgements}
We are greatful to S. Charlot for kindly providing the predictions for the SF line-emission models.
EB, FB, and GC acknowledge financial support from INAF under the Large Grant 2022 ``The metal circle: a new sharp view of the baryon cycle up to Cosmic Dawn with the latest generation IFU facilities'' and the GO grant 2024 ``A JWST/MIRI MIRACLE: Mid-IR Activity of Circumnuclear Line Emission''.
IL, FB, and AM acknowledge support from PRIN-MUR project “PROMETEUS”  financed by the European Union -  Next Generation EU, Mission 4 Component 1 CUP B53D23004750006 and C53D2300080006.
EB acknowledges INAF funding through the “Ricerca Fondamentale 2024” program (mini-grant 1.05.24.07.01). 
AM, MG, IL and CM acknowledge INAF funding through the “Ricerca Fondamentale 2023” program (mini-grant 1.05.23.04.01).
FS acknowledges financial support from the PRIN MUR 2022 2022TKPB2P - BIG-z, Ricerca Fondamentale INAF 2023 Data Analysis grant 1.05.23.03.04 ``ARCHIE ARchive Cosmic HI \& ISM  Evolution'', Ricerca Fondamentale INAF 2024 under project 1.05.24.07.01 MINI-GRANTS RSN1 "ECHOS", Bando Finanziamento ASI CI-UCO-DSR-2022-43, CUP C93C25004260005, project ``IBISCO: feedback and obscuration in local AGN''.
G.S. acknowledges the project ASI-Astrobiologia 2023 MIGLIORA (“Modeling Chemical Complexity”, F83C23000800005), the INAF-GO 2023 fundings PROTOSKA (“Exploiting ALMA data to study planet forming disks: preparing the advent of SKA”, C13C23000770005), the INAF Minigrant 2023 TRIESTE (“TRacing the chemIcal hEritage of our originS: from proTostars to planEts”; PI: G. Sabatini) and financial support under the National Recovery and Resilience Plan (NRRP), Mission 4, Component 2, Investment 1.1, Call for tender No. 104 published on 2.2.2022 by the Italian Ministry of University and Research (MUR), funded by the European Union – NextGenerationEU-Project Title 2022JC2Y93 Chemical Origins: linking the fossil composition of the Solar System with the chemistry of protoplanetary disks – CUP J53D23001600006 – Grant Assignment Decree No. 962 adopted on 30.06.2023 by the Italian Ministry of Ministry of University and Research (MUR). 
SC and GV acknowledge support by European Union’s HE ERC Starting Grant No. 101040227 - WINGS. JF acknowledges financial support from CONAHCyT, project number CF-2023-G100, and UNAM-DGAPA-PAPIIT IN111620 grant, Mexico.
AVG acknowledges support from the Spanish grant PID2022-138560NB-I00, funded by MCIN/AEI/10.13039/501100011033/FEDER, EU.
MM is thankful for support from the European Space Agency (ESA). EH gratefully acknowledges the hospitality of the IAC, where part of this work was carried out during a long research visit.
This work is based on observations made with the NASA/ESA/CSA James Webb Space Telescope. The data were obtained from the Mikulski Archive for Space Telescopes at the Space Telescope Science Institute, which is operated by the Association of Universities for Research in Astronomy, Inc., under NASA contract NAS 5-03127 for JWST. The specific observations analyzed can be accessed via doi: \url{https://doi.org/10.17909/b4w1-hk44}. 
These observations are associated with program $\#$6138. The authors acknowledge the team led by coPIs C. Marconcini and A. Feltre for developing their observing program with a zero-exclusive-access period. 
This paper makes use of the following ALMA data: ADS/JAO.ALMA$\#$2016.1.00296.S. ALMA is a partnership of ESO (representing its member states), NSF (USA) and NINS (Japan), together with NRC (Canada), NSTC and ASIAA (Taiwan), and KASI (Republic of Korea), in cooperation with the Republic of Chile. The Joint ALMA Observatory is operated by ESO, AUI/NRAO and NAOJ. 

\end{acknowledgements}

\bibliographystyle{aa}
\bibliography{aa56352-25}

\clearpage

\begin{appendix}

\section{Observations and data reduction}\label{app observation and data reduction}
In this appendix, we present the observations and data reduction for MIRI MRS, MUSE, and ALMA data.

\begin{figure*}[!t]
    \centering
    \includegraphics[width=0.8\linewidth]{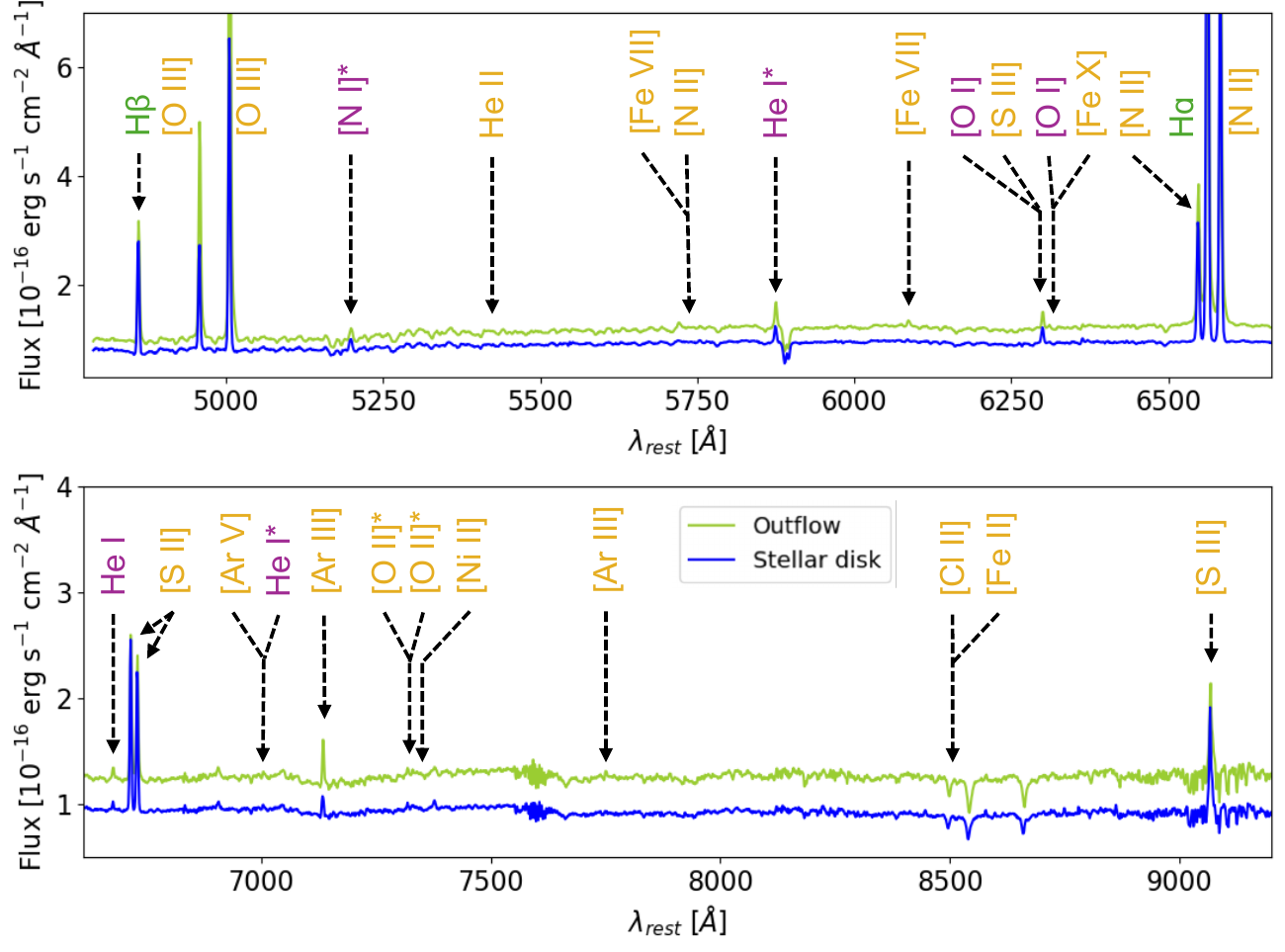}
    \caption{Integrated spectra of NGC 1365 from MUSE data. The blue and green curves represent the integrated spectra extracted with a 0.7\arcsec radius aperture from regions dominated by the stellar disk and outflow emission, respectively (see Fig.~\ref{fig: comparison fov}, lower right panel).
    Detected emission lines are marked with vertical lines: ionized gas emission lines in yellow, atomic lines in violet, and H I recombination lines in green. An asterisk indicates species corresponding to blends of two or three emission lines.}
    \label{fig: app MUSE spectra}
\end{figure*}

\subsection{JWST/MIRI}\label{app: data MIRI}
We refer to \cite{MIRACLE_NGC424} for a detailed description of the different steps performed during the data reduction and briefly summarize them here. We downloaded the uncalibrated science and background observations through the Barbara A. Mikulski Archive for Space Telescopes (MAST) portal\footnote{\url{https://www.vla.nrao.edu/astro/nvas/avla.shtml}}. The data reduction process was performed using the JWST Science Calibration Pipeline \citep{Bushouse2022_pipeline} version $1.16.0$ with JWST Calibration Reference Data System context file 1298. We applied all three stages of the pipeline processing, which include \texttt{CALWEBB\_DETECTOR1}, \texttt{CALWEBB\_SPEC2}, and \texttt{CALWEBB\_SPEC3} (see \citealt{Argyriou2020,Argyriou2023,Morrison2023,Patapis2024}). For \ngc, we subtracted the background emission from the 2D science images using a background frame generated from our dedicated background observations. Moreover, we performed two residual fringe corrections, both in stage 2 and 3. Additionally, to improve the final datacubes, we adopted the Exponential Modified-Shepard Method (EMSM) weighting function, which has proven to be more efficient in reducing the drizzling effect \citep[][]{Law2023}. 

Finally, we applied an astrometric correction to all the datacubes with different $\Delta$RA and $\Delta$Dec for each band, using the ALMA astrometry as reference (details in Appendix~\ref{app. MIRI corrections}). 

\subsection{MUSE Wide Field Mode}\label{app: data MUSE}
The data were retrieved from the ESO archive\footnote{\url{https://archive.eso.org/cms.html}} and were already processed automatically using the standard MUSE pipeline (v1.6), with an average PSF FWHM of 0.8\arcsec.
The final datacube covers a
FoV of 64\arcsec $\times$ 63.4\arcsec, with a pixel size of 0.2\arcsec pixel$^{-1}$. The MUSE FoV covers approximately the central 5.3 $\times$ 5.3 kpc$^2$ of NGC 1365. MUSE resolving power in Wide Field Mode (WFM) varies from $R\sim$ 1770 at 4800 \AA to 3590 at 9300 \AA.
Finally, we apply a relative astrometric correction, registering the position of the continuum peak, measured in the line-free 7660-8000 \AA wavelength range, to the nucleus position from the ALMA data (see Appendix~\ref{app:ALMA_data}), as done for JWST/MIRI data (see Appendix~\ref{app: data MIRI}). The applied corrections were $\Delta$RA and $\Delta$Dec shifts of (-0.78\arcsec, +0.22\arcsec).

The MUSE integrated spectra of \ngc extracted from the 0.7\arcsec radius regions shown in the lower left panel of Fig.~\ref{fig: comparison fov} are reported in Fig. \ref{fig: app MUSE spectra}. 
The green (blue) spectrum is extracted from a region where the outflow (stellar disk) emission is predominant.
We detect approximately 30 emission lines (see Table \ref{tab:homerun emission_lines 1}) tracing different gas phases, including recombination lines, atomic lines, and lines from ionized gas.

\subsection{ALMA }\label{app:ALMA_data}
The data include observations with two different configurations with angular resolutions in FWHM of $\sim0.30$\arcsec and $\sim0.07$\arcsec, respectively. In this work, we use the observations with resolution $\sim0.30$\arcsec corresponding to 30 pc, which better match the MIRI MRS spatial resolution and provide a maximum recoverable scale of 2.9\arcsec. 
We retrieved the calibrated measurement sets using the dedicated service provided by the European ALMA Regional Center \citep{Hatzimi_2015}. Imaging was performed using the Common Astronomy Software Applications ({\tt CASA}) v6.1.1 \citep[][]{CASATeam2022}. For the CO(3–2) spectral window ($342.99-344.86$~GHz), we subtracted a constant continuum level estimated from the emission line-free channels in the \textit{uv} plane. The data were cleaned using the \texttt{tclean} task in {\tt CASA}, with briggs weighting scheme with a robust parameter of 0.5, which resulted in a beam with a FWHM of 0.35\arcsec $\times$ 0.31\arcsec (beam PA = -71$^{\circ}$), corresponding to $\sim$37 pc in physical scale. 
The final datacube has a spectral channel width of $\sim$8 km s$^{-1}$, a pixel size of 0.05\arcsec, a FoV of 25\arcsec in diameter, and a root mean square of 1.5 mJy~beam$^{-1}$ per channel.

Finally, we fit the central $2\arcsec\times2\arcsec$ of the CO(3-2) flux map (see Appendix~\ref{app ALMA fitting}) with a Gaussian profile to determine the coordinates of the nucleus, taking advantage of the high astrometric accuracy provided by ALMA. We find the nuclear position to be RA = 03h33m36.37s and Dec = -36d08m25.37s, in good agreement with the values reported by \citet{Liu2023}. These coordinates are adopted throughout the rest of the paper and used to register the astrometry of MIRI and MUSE observations (see Appendices \ref{app: data MIRI} and \ref{app: data MUSE}, respectively).

\section{MIRI datacubes corrections}\label{app. MIRI corrections}
For the astrometric registration of the 12 datacubes, we applied separate corrections for each band (SHORT, MEDIUM, and LONG), as each observing setup corresponds to a different grating configuration of the MRS. As is discussed in Appendix \ref{app: data MIRI}, the observations are centered on the SE outflow cone rather than the AGN. This positioning causes the source to fall near the edge, or even slightly outside, of the FoV in the Ch1 datacubes (see Fig.~\ref{fig: comparison fov}). To ensure accurate alignment across the dataset, we computed the astrometric corrections using Ch3, where the source is well within the FoV and the spatial resolution remains high, providing a reliable reference frame for registration.
We collapsed the datacubes along the spectral dimension and fit the PSFs with a 2D Gaussian model. Then, we aligned the datacubes to match the PSF peaks with the AGN coordinates provided ALMA data (see Appendix \ref{app:ALMA_data}). The applied corrections were $\Delta$RA and $\Delta$Dec shifts of (+0.0086\arcsec, -0.2655\arcsec), (+0.0256\arcsec, -0.2713\arcsec), and (-0.0300\arcsec, -0.299\arcsec) for the SHORT, MEDIUM, and LONG channels, respectively.

\begin{figure}[t!]
    \centering
    \includegraphics[width=1\linewidth]{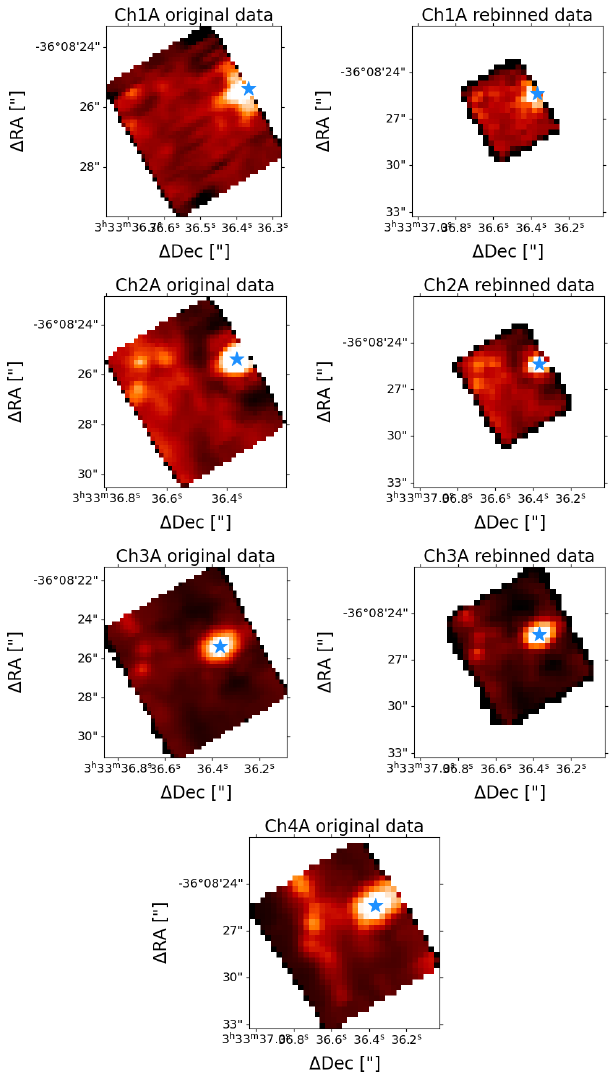}
    \caption{Examples of the rebinning procedure for datacubes in the A (SHORT) band. The first three rows compare the original collapsed data with native pixel size and FoV (left) to the rebinned data matched to the wider FoV of Ch4 (right). The bottom row shows the original Ch4 data.
    The star marks the nucleus position.}
    \label{fig: reprojected maps}
\end{figure}
To compare flux maps of emission lines from different channels it is crucial to account for the different pixel size (see Appendix \ref{app: data MIRI}). To address this, we rebinned each channel to match the pixel size of Ch4, which has the worst resolution (0.35\arcsec). We used the \texttt{reproject exact} function to ensure flux conservation during the resampling process.
In Fig.~\ref{fig: reprojected maps} we show the collapsed datacubes before and after the spatial rebinning. The relative flux conservation error is below the 1$\%$, so negligible respect the original flux.

\begin{figure*}[!t]
    \centering
    \includegraphics[width=.99\linewidth]{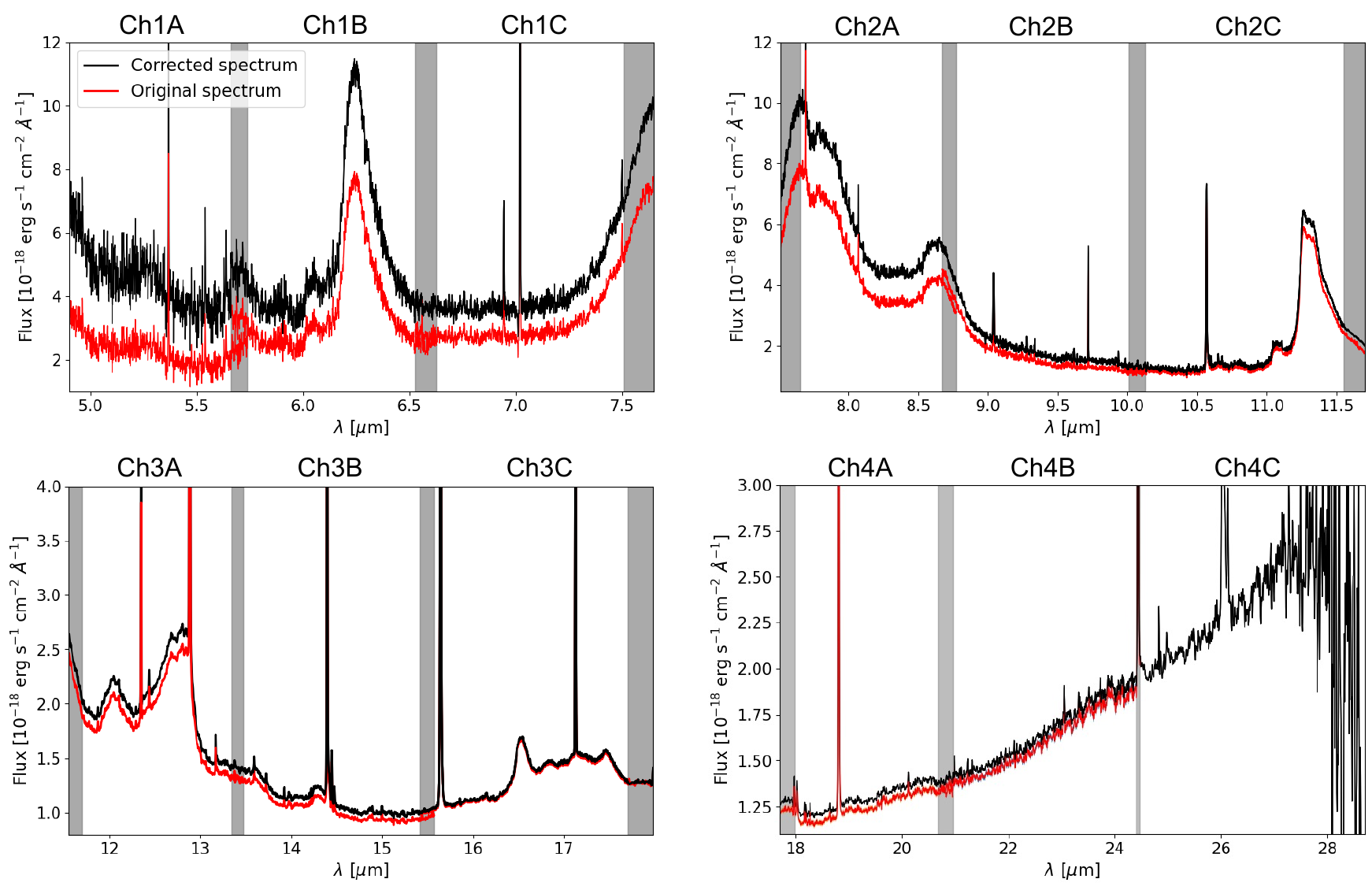}
    \caption{Comparison per channel between the original data in red and the stitched data in black. gray regions indicate overlapping spectral channels.}
    \label{fig: app stitching}
\end{figure*}

MIRI data are known to exhibit flux discontinuities between different sub-channels. This effect could be amplified after the rebinning for the different pixel size of data. Therefore, it is essential to apply a scaling factor to the fluxes when constructing a single spectrum, in addition to account for differences in spectral sampling. To address this, we implemented a stitching procedure between each pair of datacubes by rescaling the flux of the bluer datacube to match the median flux value of the overlapping spectral channels of the redder datacube.
We chose to rescale to the reddest datacube (Ch4 LONG) as it has the largest FoV, ensuring consistent scaling in every spaxel. This process yielded a scaling factor matrix for each pair of datacubes, which was then applied to correct all fluxes derived from our data.
In Fig.~\ref{fig: app stitching} an example of the total spectrum before and after the stitching procedure.

As shown in \citet{Law2023}, the average FWHM of the MIRI/MRS PSF varies across the four channels spanning the range FWHM = 0.4\arcsec, 0.5\arcsec, 0.6\arcsec, and 0.9\arcsec, in Ch1, Ch2, Ch3, and Ch4, respectively. This variation must be taken into account when computing ratios between emission lines at different wavelengths. To ensure consistent spatial resolution, we firstly downloaded the MIRI MRS PSF model cube using the \textit{WebbPSF} tool v.1.4.0 \citep{Perrin2014}, which provide a model cube spatially oversampled of a factor of 4 with respect to the observations (i.e., 0.05 \arcsec/pixel for Ch3).
Then we deconvolved each flux map with the PSF corresponding to the wavelength of the emission line, and convolved it with the PSF at the reddest wavelength among the lines involved in the ratios, using the PSF model cubes from \textit{WebbPSF}. The deconvolution procedure is composed by a wavelength-dependent alignment, rotation and flux normalization between the PSFs and our flux maps, following a similar methodology of \cite{MIRACLE_NGC424}.

\section{Data analysis}\label{app data analysis}
In this appendix, we describe the general methodology employed for the emission line analysis of data from MUSE, ALMA, and MIRI MRS. In particular, we analyzed the MUSE and MIRI data to subtract the continuum and fit the emission lines in the optical and the mid-IR band, using a multi-Gaussian fitting procedure.

\subsection{JWST/MIRI MRS emission line fitting} \label{app miri emission line fitting}
This appendix provides further details on the customized python routine we employed to analyze the MIRI MRS datacubes. We followed a three-step procedure: first, we applied spatial smoothing to enhance the S/N; second, we subtracted the local continuum around the brightest emission lines; finally, we performed spaxel-by-spaxel Gaussian fitting of the residual emission.
First, we applied a Gaussian spatial smoothing with a kernel size of 1 pixel to every spatial plane of the datacube. 
This step improves the S/N of faint extended features, at the cost of a moderate degradation in spatial resolution. Specifically, the effective PSF is broadened by approximately 40$\%$ across all wavelengths, due to the fact that the pixel size is comparable to or slightly smaller than the intrinsic PSF $\sigma$ at each wavelength. Nevertheless, the resulting resolution remains sufficient for our spatially resolved analysis, as it still allows us to distinguish the relevant morphological and kinematic structures in the data.

\begin{table}[h!]
\caption{List of the fit emission lines in the MIRI spectra}
\centering
\begin{tabular}{lcccc}
\toprule\toprule
Emission line         & $\lambda^{(1)}$  &  IP$^{(1)}$   & Ch & Flux  \\
                      & [$\mu$m]           & [eV]    &         & [$10^{-14}$ erg s$^{-1}$ cm$^{-2}$] \\
\hline
\text{[\ion{Mg}{V}]}         & 5.61               & 109     & 1A      & 1.75~$\pm$~0.05 \\
\text{Pf$\alpha$}     & 7.46               & -       & 1C      & 0.962~$\pm$~0.005 \\
\text{[\ion{Ne}{VI}]}        & 7.65               & 126     & 2A      & 8.44~$\pm$~0.09 \\
\text{[\ion{Ar}{V}]}         & 7.90               & 60      & 2A      & 1.4~$\pm$~0.1 \\
H$_2$ 0-0 S(4)        & 8.03               & -       & 2A      & 2.06~$\pm$~0.04 \\
\text{[\ion{Ar}{III}]}       & 8.99               & 28      & 2B      & 3.4~$\pm$~0.2 \\
H$_2$ 0-0 S(3)        & 9.66               & -       & 2B      & 4.6~$\pm$~0.6 \\
\text{[\ion{S}{IV}]}         & 10.51              & 35      & 2C      & 24.0~$\pm$~0.4 \\
H$_2$ 0-0 S(2)        & 12.28              & -       & 3A      & 7.5~$\pm$~0.1 \\
\text{Hu$\alpha$}     & 12.37              & -       & 3A      & 0.760~$\pm$~0.002 \\
\text{[\ion{Ne}{II}]}        & 12.81              & 23      & 3A      & 102.6~$\pm$~0.4 \\
\text{[\ion{Ar}{V}]}         & 13.10              & 60      & 3A      & 0.702~$\pm$~0.004 \\
\text{[\ion{Ne}{V}]}         & 14.32              & 97      & 3B      & 22.2~$\pm$~0.4 \\
\text{[\ion{Cl}{II}]}        & 14.37              & 13      & 3B      & 2.02~$\pm$~0.03 \\
\text{[\ion{Ne}{III}]}       & 15.56              & 41      & 3C      & 56.6~$\pm$~0.4 \\
H$_2$ 0-0 S(1)        & 17.04              & -       & 3C      & 16.3~$\pm$~0.4 \\
\text{[\ion{S}{III}]}        & 18.71              & 23      & 4A      & 50.5~$\pm$~0.2 \\
\text{[\ion{Ne}{V}]}         & 24.32              & 97      & 4C      & 26.70~$\pm$~0.04 \\
\text{[\ion{O}{IV}]}         & 25.89              & 55      & 4C      & 53.0~$\pm$~2.0 \\
\bottomrule
\label{tab: table list emission lines}
\end{tabular}
\tablefoot{From left to right: the name of the emission line, the corresponding rest-frame wavelength, the IP, the MIRI channel where it is observed, and the integrated flux within the FoV of the respective channel, expressed in units of $10^{-14}$ erg s$^{-1}$ cm$^{-2}$.}\\
\tablebib{(1)~\hyperlink{https://www.mpe.mpg.de/ir/ISO/linelists/FSlines.html}{ISO Spectrometer Data Centre.}}
\end{table}

We then performed the emission line fitting spaxel-by-spaxel, fitting each line in Table \ref{tab: table list emission lines} independently. Prior to this, we modeled and fit the local continuum selecting wavelength windows on either side of the line, adjusting the range based on the FWHM of each emission line and carefully avoiding other spectral absorption or emission features. Given the short spectral window of $\sim$0.1~$\mu$m we considered, within which the continuum is featureless, we fit the continuum with a straight line.
Each emission line is modeled with both a single and a double Gaussian component, and then we applied a Kolmogorov-Smirnov (KS) test on the residuals to choose the optimal number of Gaussian components in each spaxel, adopting a threshold p-value of 0.7  (see \citealp{Marasco_2020, Tozzi2021, Ceci_2025} for details).

In the mid-IR, dust attenuation is much lower than at optical wavelengths, but it does vary with wavelength and affect emission lines differently, potentially impacting specific line ratios. Unfortunately, in our data the S/N of the Pf$\alpha$ and Hu$\alpha$ (i.e., H~I~(6−5) at 7.46~$\mu$m and H~I~(7−6) at 12.37~$\mu$m, respectively) is too low to derive a reliable extinction map directly from the mid-IR spectra. 
To test the potential impact of dust attenuation, we applied an extinction correction to the mid-IR emission lines using the G23 \citep{Gordon09,Gordon21,Gordon+2023,Fitzpatrick19,Decleir22} attenuation curve, from the \textit{dust$\_$extinction} package\footnote{https://github.com/karllark/dust$\_$extinction} \citep{Gordon24_dustextinction_package}, and adopting the maximum $A_V$ value ($\sim$3) derived from our best-fit photoionization model (see Section~\ref{sec HOMERUN fitting results}). However, we find that the resulting variations in the mid-IR line ratios are within the errors, including those involving emission lines at distant wavelengths (e.g., \nevB/\nevA, see Section~\ref{sec density}).

\subsection{MUSE WFM emission line fitting} \label{app MUSE emission line fitting}

MUSE data were analyzed as already described in previous works of the MAGNUM program \citep{Venturi_2018, Mingozzi2019}. Briefly, after a preliminary Voronoi tessellation (\citealp{Cappellari2003}) to achieve an average S/N per wavelength channel of at least 70 on the continuum, we applied the penalized PiXel-Fitting (pPXF,  \citealp{Cappellari2004}) code to measure the optical emission lines listed in Table 
\ref{tab:homerun emission_lines 1}.
Using the penalized PiXel-Fitting (pPXF, \citealp{Cappellari2004}) code, we fit simultaneously the continuum and gas emission in the wavelength range 4800-9200 \AA, convolving a linear combination of stellar templates (E-MILES; for details see \citealp{Vazdekis2016}) with a Gaussian velocity distribution.
Each emission line was fit twice, using one and two Gaussian components, and the KS test was applied spaxel-by-spaxel to determine the minimum number of components required to adequately reproduce the line profiles (see Appendix~\ref{app miri emission line fitting}).
Given that \ngc hosts a Seyfert 1.8 nucleus, we included an additional spectral component when fitting the bins within a radius of 10 spaxels from the nuclear position to account for the spatially unresolved broad-line region (BLR). The BLR model consists of two independent components: a broad ($\sigma$>1000 km s$^{-1}$) Gaussian component for the broad Balmer lines (i.e., H$\alpha$ and H$\beta$) and a template of broad [\ion{Fe}{II}] emission features by \cite{BorosonGreen1992}.
The nuclear model spectrum was obtained by fitting the integrated spectrum of the central spaxels and, in the subsequent spaxel-by-spaxel fitting, it was kept fixed in shape and velocity while being allowed to scale in flux. 
We then subtracted the stellar and nuclear emission from the original unbinned data cube by rescaling the fit continuum - assumed constant within each Voronoi bin - to match the observed continuum flux in each spaxel, before performing the subtraction.

We obtained continuum-subtracted datacubes containing the gas emission lines only. Then, we repeated the fitting procedure applied for MIRI JWST data (see previous section). In this case, we used up to three Gaussian components to reproduce the complex line profiles and adopted a p-value of 0.9.

Finally, we corrected the optical emission-line fluxes for dust attenuation using the Balmer decrement (i.e., H$\alpha$/H$\beta$), assuming a case-B hydrogen recombination value of 2.86 for this ratio \citep[for an electron temperature and density of T$_e$~=~$10^4$~K and n$_e$~=~$10^2$~cm$^{-3}$, respectively;][]{Osterbrock2006} and adopting the G23 attenuation law. The results, presented in Appendix~\ref{app extinction&density from muse}, are consistent with previous estimates by \citet{Venturi_2018} and \citet{Mingozzi2019}. As shown in Fig.~\ref{fig: app extinction&density from muse}, the dust attenuation peaks at A$_V$ = 3.7 in the circumnuclear ring, where young stellar clusters are embedded \citep{Liu2023, Schinnerer2023}. High attenuation values are also found at the nuclear position, while we do not find any correlation of A$_V$ with the outflow morphology, indicative of its low dust content.

\subsection{ALMA emission line fitting}\label{app ALMA fitting}
We use the CO(3-2) ALMA observations to derive the kinematics of the cold molecular gas.
The continuum was subtracted in the $uv$-plane during the data reduction (see Appendix~\ref{app:ALMA_data}).
We fit the CO(3-2) line profile in each spaxel using a single Gaussian profile, which is sufficient to reproduce the symmetric line profile. The resulting moment maps of the cold molecular gas are reported in the bottom panels of Fig.~\ref{fig: moment maps}.

\section{Extinction and electron density from MUSE}\label{app extinction&density from muse}
\begin{figure}
    \centering
    \includegraphics[width=1\linewidth]{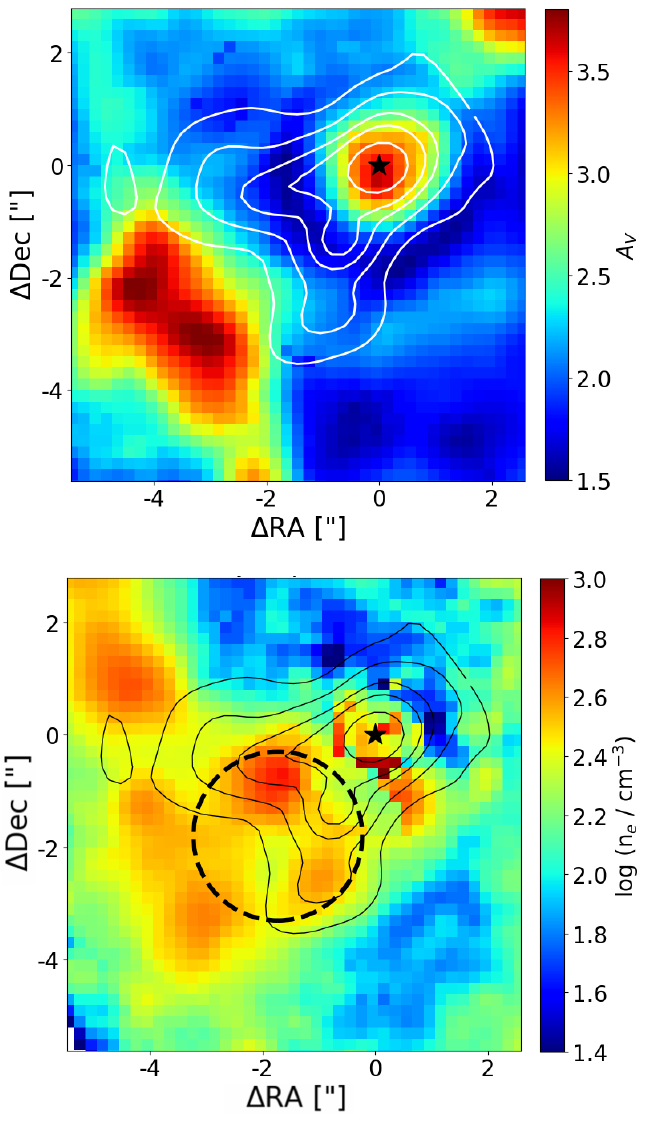}
    \caption{Spatially resolved maps of visual extinction and electron density in the circumnuclear region of \ngc obtained from the MUSE data, zoomed in the MIRI FoV.
    \textit{Top panel:} visual extinction computed from the Balmer decrement.
    \textit{Bottom panel:} ionized gas electron density obtained from the \siia/\siib line ratio. Contours represent arbitrary flux levels of \nevA. The black circle marks the 1.5\arcsec radius aperture used to extract integrated fluxes for the HOMERUN code (see Section \ref{sec: homerun}). The star indicates the nucleus position.}
    \label{fig: app extinction&density from muse}
\end{figure}
In this appendix, we present the spatially resolved maps of visual extinction and electron density in the circumnuclear region of \ngc, derived from the MUSE data and zoomed-in FoV matching the MIRI one. In the top panel of Fig. \ref{fig: app extinction&density from muse}, we show the visual extinction computed from the Balmer decrement (H$\alpha$/H$\beta$), adopting the G23 extinction law. High visual extinction values of A$_V \sim$~3.7 mag are found at the nuclear position and in the SE part of the FoV, corresponding to the circumnuclear ring, as also observed by \citet{Venturi_2018} and \citet{Mingozzi2019}.

Moreover, the bottom panel of Fig.~\ref{fig: app extinction&density from muse} shows the electron density of the ionized gas, estimated from the total intensity of all the Gaussian components of the \siiALL doublet. In particular, we use the \siia/\siib diagnostic line ratio \citep{Osterbrock2006}, computed over the same FoV of the Ch4. 
We assume a typical value for the temperature of ionized gas of T$_e$ = 10$^4$ K and obtain the density map reported in Fig.~\ref{fig: app extinction&density from muse}. We observe that the higher values of n$_e$ are following the outflow morphology, as traced by \nevA contours flux.
In the main text, we use as average density the median value and the standard deviation of n$_e$ = (230 $\pm$ 120) cm$^{-3}$ measured across the reported FoV.

\section{Density comparison with IRS observations} \label{app dudik LDL}
In this appendix, we compare our spatially resolved JWST/MIRI density measurements with those obtained from integrated spectra by \citet{Dudik+07}.
\cite{Dudik+07} used the Infrared Spectrograph (IRS) on board {\it Spitzer} to compute the [\ion{Ne}{V}] ratio in a sample of local AGNs.  
They found that Type 2 AGNs show higher line ratios with respect to Type 1, with all Type 2 sources lying above the theoretical LDL threshold, implying log(n$_e$/cm$^{-3}$)$~\le~$2. In particular, for \ngc they computed a \nevB/\nevA ratio of 2.4$\pm$0.6, which is larger than our estimation of 1.0$\pm$0.2 obtained as the ratio between the two integrated fluxes in the MIRI FoV.
They argued that the [\ion{Ne}{V}] emitting region likely originates inside the nuclear torus and that differential IR extinction due to dust in the obscuring torus may be responsible for the ratio above the LDL.
Exploiting the spatially resolved [\ion{Ne}{V}] emission in our data and the density values in the nuclear region, we can rule out this interpretation. However, given the presence of some LDL regions in Fig. \ref{fig: density}, it is possible that their finding is the result of spatial integration over a wide slit, and thus including such regions in the measurement.  
A more plausible reason is an observational effect. Indeed, in \cite{Dudik+07} the [\ion{Ne}{V}] fluxes were obtained using the short-wavelength, high-resolution (SH; 4.7\arcsec$\times$11.3\arcsec, $\lambda$~=~9.9-19.6 $\mu$m) and long-wavelength, high-resolution (LH; 11.1\arcsec$\times$22.3\arcsec, $\lambda$~=~18.7-37.2 $\mu$m) modules of IRS. As it is clear from the moment maps of the high-IP species in the online material, the extension of [\ion{Ne}{V}] emission is detected up to 8\arcsec from the nucleus. This means that the SH module surely lost part of the \nevA flux, leading to an artificially enhancement of the \nevB/\nevA line ratio.

\section{HOMERUN emission line fluxes}\label{app homerun results}

\begin{table*}[htbp]
\centering
\caption{Optical emission lines fit by the HOMERUN code}
\label{tab:homerun emission_lines 1}
\normalsize
\begin{tabular}{lccccccccc}
\toprule\toprule
Emission line & $\lambda$ \,[\textnormal{$\AA$}] & Obs. Luminosity & Model & Mod/Obs & AGN frac. & SF frac. & AGN dered. & SF dered. \\\\
~~~~~~~~(1) & (2) & (3) & (4) & (5) & (6) & (7) & (8) & (9) \\
\midrule
H$\beta$ & 4861 & $100.00 \pm 10.00$ & $90.34$ & $0.90$ & $0.76$ & $0.24$ & $302.27$ & $702.62$ \\
{[O~III]} & 5007 & $465.99 \pm 46.60$ & $500.05$ & $1.07$ & $0.97$ & $0.04$ & $2022.68$ & $500.44$ \\
{[N~I]$^*$} & 5199 & $7.51 \pm 0.75$ & $7.53$ & $1.00$ & $1.00$ & $0.00$ & $29.67$ & $0.00$ \\
He~II & 5411 & $3.04 \pm 0.32$ & $3.08$ & $1.01$ & $1.00$ & $0.00$ & $11.39$ & $0.00$ \\
{[Fe~VII]} & 5721 & $4.77 \pm 0.95$ & $3.64$ & $0.76$ & $1.00$ & $0.00$ & $12.35$ & $0.00$ \\
{[N~II]} & 5755 & $2.41 \pm 0.42$ & $2.51$ & $1.04$ & $0.64$ & $0.36$ & $5.37$ & $15.75$ \\
{[He~I]$^*$} & 5876 & $19.75 \pm 1.97$ & $19.45$ & $0.99$ & $0.69$ & $0.31$ & $43.90$ & $95.28$ \\
{[Fe~VII]} & 6087 & $6.30 \pm 0.63$ & $5.98$ & $0.95$ & $1.00$ & $0.00$ & $18.46$ & $0.00$ \\
{[O~I]} & 6300 & $13.55 \pm 1.35$ & $13.68$ & $1.01$ & $0.75$ & $0.25$ & $30.01$ & $43.22$ \\
{[S~III]} & 6312 & $1.43 \pm 0.42$ & $1.90$ & $1.33$ & $0.82$ & $0.18$ & $4.55$ & $4.34$ \\
{[O~I]} & 6364 & $4.60 \pm 0.46$ & $4.48$ & $0.98$ & $0.75$ & $0.25$ & $9.67$ & $13.81$ \\
{[Fe~X]} & 6375 & $1.19 \pm 0.28$ & $1.22$ & $1.03$ & $1.00$ & $0.00$ & $3.53$ & $0.00$ \\
H$\alpha$ & 6563 & $622.41 \pm 62.24$ & $660.61$ & $1.06$ & $0.72$ & $0.28$ & $1322.36$ & $2000.05$ \\
{[N~II]} & 6583 & $378.71 \pm 37.87$ & $369.45$ & $0.98$ & $0.31$ & $0.69$ & $318.65$ & $2748.56$ \\
He~I & 6678 & $7.00 \pm 0.70$ & $6.75$ & $0.96$ & $0.62$ & $0.38$ & $11.25$ & $26.74$ \\
{[S~II]} & 6716 & $68.33 \pm 6.83$ & $71.35$ & $1.04$ & $0.26$ & $0.74$ & $49.67$ & $535.85$ \\
{[S~II]} & 6731 & $63.03 \pm 6.30$ & $52.86$ & $0.84$ & $0.27$ & $0.73$ & $38.19$ & $388.58$ \\
{[Ar~V]} & 7006 & $2.07 \pm 0.46$ & $1.72$  & $0.83$ & $1.00$ & $0.00$ & $4.36$ & $0.00$ \\
{[He~I]$^*$} & 7065 & $2.56 \pm 25.62$ & $12.78$ & $4.98$ & $0.79$ & $0.21$ & $25.34$ & $22.95$ \\
{[Ar~III]} & 7136 & $17.56 \pm 1.76$ & $16.34$ & $0.93$ & $0.61$ & $0.40$ & $24.44$ & $54.04$ \\
Ca~II & 7291 & $< 1.10$ & $0.00$ & $0.00$ & $0.00$ & $0.00$ & $0.00$ & $0.00$ \\
{[O~II]$^*$} & 7323 & $3.11\pm 0.68$ & $2.87$ & $0.92$ & $0.60$ & $0.40$ & $4.10$  & $8.94$ \\
{[O~II]$^*$} & 7332 & $2.41\pm 0.76$ & $2.35$ & $0.97$ & $0.60$ & $0.40$ & $3.34$  & $7.32$ \\
{[Ni~II]} & 7378 & $2.86 \pm 0.43$ & $2.75$ & $0.96$ & $1.00$ & $0.00$ & $6.52$ & $0.00$ \\
{[Ar~III]} & 7751 & $4.54 \pm 0.45$ & $4.56$ & $1.00$ & $0.57$ & $0.43$ & $5.80$ & $12.83$ \\
{[Cl~II]} & 8579 & $2.48 \pm 0.68$ & $3.52$ & $1.42$ & $0.00$ & $1.00$ & $0.00$ & $17.57$ \\
{[Fe~II]} & 8617 & $0.85 \pm 0.67$ & $0.74$ & $0.88$ & $0.51$ & $0.49$ & $0.75$ & $1.79$ \\
{[S~III]} & 9069 & $61.72 \pm 6.20$ & $62.49$ & $1.01$ & $0.49$ & $0.52$ & $56.72$ & $140.12$ \\
\bottomrule
\end{tabular}
\tablefoot{
(1) Name of the emission line. Lines marked with an asterisk correspond to transitions that in the \textsc{CLOUDY} framework are treated as blended multiplets.
(2) Rest-frame wavelength of the emission line. 
(3) Observed luminosity scaled to an H$\beta$ flux of 100. For some emission lines, error is set to 1000\% to reflect highly uncertain measurements, marginal line detections dominated by noise, or to exclude those lines from the fit. 
(4) Line luminosity predicted by the model. 
(5) Ratio between the model and observed luminosity. 
(6) Fractional contribution of the AGN component to the observed line luminosity. 
(7) Fractional contribution of the SF component to the observed line luminosity. 
(8) Dereddened model luminosity of the AGN component normalized as in column 3. 
(9) Dereddened model luminosity of the SF component normalized as in column 3.
}
\end{table*}

\begin{table*}[htbp]
\centering
\caption{Mid-IR emission lines fit by the HOMERUN code}
\label{tab:homerun emission_lines 2}
\normalsize
\begin{tabular}{lcccccccc}
\toprule\toprule
Emission line & $\lambda$ \,[\textnormal{$\mu$m}] & Obs. Luminosity & Model & Mod/Obs & AGN frac. & SF frac. & AGN dered. & SF dered. \\
~~~~~~~~(1) & (2) & (3) & (4) & (5) & (6) & (7) & (8) & (9) \\
\midrule
{[Fe~II]} & 5.34 & $140.58 \pm 14.06$ & $129.03$ & $0.92$ & $0.16$ & $0.84$ & $21.24$ & $116.24$ \\
{[Fe~VIII]} & 5.44 & $14.01 \pm 140.07$ & $6.92$ & $0.49$ & $1.00$ & $0.00$ & $7.12$ & $0.00$ \\
{[Mg~VII]} & 5.50 & $17.20 \pm 2.79$ & $18.04$ & $1.05$ & $1.00$ & $0.00$ & $18.57$ & $0.00$ \\
{[Mg~V]} & 5.61 & $95.37 \pm 9.54$ & $82.35$ & $0.86$ & $1.00$ & $0.00$ & $84.74$ & $0.00$ \\
{[Ni~II]} & 6.64 & $8.30 \pm 0.99$ & $8.57$ & $1.03$ & $1.00$ & $0.00$ & $8.82$ & $0.00$ \\
{[Fe~II]} & 6.72 & $6.28 \pm 62.75$ & $9.37$ & $1.50$ & $0.16$ & $0.84$ & $1.54$ & $8.43$ \\
{[Ar~II]} & 6.98 & $384.47 \pm 3844.69$ & $72.16$ & $0.19$ & $0.04$ & $0.96$ & $2.97$ & $74.35$ \\
{[Na~III]} & 7.32 & $16.93 \pm 1.69$ & $19.32$ & $1.14$ & $1.00$ & $0.00$ & $19.96$ & $0.00$ \\
Pa$\alpha$ & 7.46 & $25.02 \pm 2.50$ & $24.06$ & $0.96$ & $0.36$ & $0.64$ & $9.02$ & $16.62$ \\
{[Ne~VI]} & 7.65 & $222.70 \pm 22.27$ & $235.62$ & $1.06$ & $1.00$ & $0.00$ & $244.35$ & $0.00$ \\
{[Fe~VII]} & 7.81 & $4.22 \pm 1.32$ & $6.01$ & $1.43$ & $1.00$ & $0.00$ & $6.25$ & $0.00$ \\
{[Ar~V]} & 7.90 & $8.34 \pm 1.18$ & $8.63$ & $1.04$ & $1.00$ & $0.00$ & $8.99$ & $0.00$ \\
{[Na~VI]} & 8.61 & $11.93 \pm 119.32$ & $5.41$ & $0.45$ & $1.00$ & $0.00$ & $5.74$ & $0.00$ \\
{[Ar~III]} & 8.99 & $97.69 \pm 9.77$ & $112.34$ & $1.15$ & $0.22$ & $0.78$ & $26.98$ & $104.45$ \\
{[Na~IV]} & 9.04 & $12.77 \pm 1.28$ & $12.67$ & $0.99$ & $1.00$ & $0.00$ & $13.70$ & $0.00$ \\
{[Fe~VII]} & 9.53 & $15.73 \pm 1.57$ & $15.87$ & $1.01$ & $1.00$ & $0.00$ & $17.55$ & $0.00$ \\
{[S~IV]} & 10.51 & $426.55 \pm 42.66$ & $453.92$ & $1.06$ & $0.95$ & $0.05$ & $474.43$ & $25.76$ \\
Hu$\alpha$ & 12.37 & $7.82 \pm 0.78$ & $8.60$ & $1.10$ & $0.36$ & $0.64$ & $3.24$ & $6.15$ \\
{[Ne~II]} & 12.81 & $974.91 \pm 97.49$ & $944.99$ & $0.97$ & $0.02$ & $0.98$ & $23.79$ & $1022.99$ \\
{[Ar~V]} & 13.10 & $16.32 \pm 1.63$ & $14.11$ & $0.87$ & $1.00$ & $0.00$ & $14.74$ & $0.00$ \\
{[Mg~V]} & 13.52 & $4.97 \pm 0.50$ & $5.53$ & $1.11$ & $1.00$ & $0.00$ & $5.77$ & $0.00$ \\
{[Ne~V]} & 14.32 & $389.91 \pm 38.99$ & $339.14$ & $0.87$ & $1.00$ & $0.00$ & $353.89$ & $0.00$ \\
{[Cl~II]} & 14.37 & $19.20 \pm 1.92$ & $17.49$ & $0.91$ & $0.00$ & $1.00$ & $0.00$ & $19.33$ \\
{[Na~VI]} & 14.40 & $9.54 \pm 0.95$ & $8.36$ & $0.88$ & $1.00$ & $0.00$ & $8.73$ & $0.00$ \\
{[Ne~III]} & 15.56 & $593.28 \pm 59.33$ & $543.72$ & $0.92$ & $0.71$ & $0.29$ & $401.92$ & $177.98$ \\
{[Fe~II]} & 17.94 & $10.99 \pm 1.10$ & $12.65$ & $1.15$ & $0.19$ & $0.81$ & $2.59$ & $11.53$ \\
{[S~III]} & 18.71 & $386.66 \pm 38.67$ & $360.72$ & $0.93$ & $0.24$ & $0.77$ & $89.40$ & $312.78$ \\
{[Fe~VI]} & 19.56 & $8.76 \pm 0.88$ & $8.94$ & $1.02$ & $1.00$ & $0.00$ & $9.42$ & $0.00$ \\
{[Fe~III]} & 22.93 & $17.24 \pm 1.72$ & $16.91$ & $0.98$ & $0.31$ & $0.69$ & $5.54$ & $13.03$ \\
{[Ne~V]} & 24.32 & $388.76 \pm 38.88$ & $442.87$ & $1.14$ & $1.00$ & $0.00$ & $463.94$ & $0.00$ \\
{[O~IV]} & 25.89 & $1535.40 \pm 153.54$ & $1278.60$ & $0.83$ & $1.00$ & $0.00$ & $1337.12$ & $0.01$ \\
{[Fe~II]} & 25.99 & $57.59 \pm 11.52$ & $45.21$ & $0.79$ & $0.20$ & $0.80$ & $9.43$ & $40.19$ \\
\bottomrule
\end{tabular}
\tablefoot{See Table \ref{tab:homerun emission_lines 1}.}
\end{table*}

In this appendix we report the full list of emission lines included in the HOMERUN fit, as shown in Tables~\ref{tab:homerun emission_lines 1} and \ref{tab:homerun emission_lines 2}, and discussed in Section~\ref{sec HOMERUN fitting results}.

\section{Ionized gas mass estimate}\label{app mass estimation}
In this appendix we present an analytical mass estimation for the \nevA and the \oiii emission lines, in order to compute the gas masses in the 1.5\arcsec radius aperture (shown as a dashed circle in Fig. \ref{fig: app extinction&density from muse}) and compare them with the HOMERUN estimations in Section \ref{sec Ionized gas mass from HOMERUN}.

Following \cite{Carniani2015}, we start from the luminosity of the \nevA line:
\begin{equation} \label{eq: L NeV}
    L_\text{[Ne V]} = \int_{V} f \, n_e\, n\bigr(Ne^{4+}\bigr)\, \gamma_\text{[Ne\,V]}({n_e, T_e}) dV
\end{equation}
with f the filling factor, V the volume occupied by the outflowing ionized gas, $n_e$ the electron density, $n(Ne^{4+})$ the density of Ne$^{4+}$ ions, and $\gamma_\text{[Ne\,V]}({n_e, T_e})$ the line emissivity. 
$n(Ne^{4+})$ can be written as 
\begin{equation*}\label{eq density for outflow}
    n\left(Ne^{4+}\right)\ = \left[ \frac{ n\bigr(Ne^{4+}\bigr)}{n\left(Ne\right)} \right] \left[ \frac{ n\left(Ne\right)}{n\left(H\right)} \right]    \left[ \frac{ n\left(H\right)}{n_e} \right] n_e.
\end{equation*}

We assume $n(Ne^{4+})$ $\approx$ $n(Ne)$, $n_e$ $\approx$ 1.2 $n(H)$ (i.e., a 10\% number density of He atoms with respect to H atoms), and a solar abundance of [Ne/H] $\sim$ 7.93 \citep{Asplund+2009}, to obtain $n(Ne^{4+}) \sim 7.09 \times 10^{-5} \, n_e$. 
Assuming a typical temperature (T$_e \simeq 10^4 K$), from the \nevB/\nevA line ratio we find an electron density of n$_e$ = (750$\pm$440)~cm$^{-3}$ in the 1.5" radius aperture and estimate a line emissivity of $\gamma_\text{[Ne\,V]} = 1.93 \times 10^{-20}$ erg s$^{-1}$ cm$^{3}$ making use of PyNeb \citep{Luridiana+2015}. Finally, Eq. \ref{eq: L NeV} can be rewritten as 
\begin{equation} \label{eq: L NeV 2}
    L_\text{[Ne V]} = 7.09 \times 10^{-5} f \, n_e^2 \, \gamma_\text{[Ne\,V]} \, V,
\end{equation}
assuming a constant electron density in the outflow volume.
The gas mass can be expressed as 
\begin{equation}\label{eq mass for outflow} 
    M = \int_{V} f \, \bar{m }\, n(H)dV \simeq \int_{V} f \, m_p \, n_e dV,
\end{equation}
where the average molecular weight is $\bar{m}$ = 1.27 $m_p$, if there is a 10\% fraction of He atoms, and we have taken into account that 
\begin{equation} \label{eq average mol weight}
    \bar{m}\, n(H) = \bar{m} \left[ \frac{n(H)}{n_e} \right] \, n_e = 1.27 \, m_p \, (1.2)^{-1} \, n_e \approx m_p \, n_e;
\end{equation}
therefore, $M \approx f \, m_p \, n_e \, V$. Combining this latter with Equation \ref{eq: L NeV 2}, we finally get
\begin{equation} \label{eq: M outflow NeV}
    M_\text{[Ne V]} = 3.07 \, M_\odot 
    \left( \frac{L_\text{[Ne V]}}{10^{36} \, \text{erg} \, \text{s}^{-1}} \right) 
    \left( \frac{n_e}{200 \, \text{cm}^{-3}} \right)^{-1}. 
\end{equation}

Similarly, we computed the \oiii outflow mass. Using a solar abundance of [O/H] $\sim$ 8.69 \citep{Asplund+2009} and n$_e$ = (500~$\pm$~270) cm$^{-3}$ (from the \siiALL ratio in the same aperture) - that leads to $\gamma_\text{[O\,III]} = 3.53 \times 10^{-21}$ erg s$^{-1}$ cm$^{3}$ - we have
\begin{equation} \label{eq: M outflow OIII}
    M_\text{[O III]} = 2.92 \, M_\odot 
    \left( \frac{L_\text{[O III]}}{10^{36} \, \text{erg} \, \text{s}^{-1}} \right) 
    \left( \frac{n_e}{200 \, \text{cm}^{-3}} \right)^{-1} .
\end{equation}

We corrected both the \nevA and the \oiii luminosity for dust attenuation using the Balmer decrement from the H$_\alpha$/H$_\beta$ ratio, employing the G23 attenuation law and an intrinsic ratio H$_\alpha$/H$_\beta$ = 2.86 (for T$_e$ = 10$^4$ K, \citealp{Osterbrock2006}). 
Assuming the two different densities for the optical and mid-IR species previously computed, we obtain an outflow mass of $\rm M_{[Ne V]}$ = (2.6~$\pm$~1.0)~$\times$~10$^{3}$~\msun and $\rm M_{[O III]}$ = (20~$\pm$~7)~$\times$~10$^{4}$~\msun.

\end{appendix}

\end{document}